\documentclass[journal]{vgtc}                % final (journal style)
\ifpdf%                                % if we use pdflatex
  \pdfoutput=1\relax                   % create PDFs from pdfLaTeX
  \usepackage{graphicx}                % allow us to embed graphics files
  \DeclareGraphicsExtensions{.pdf,.png,.jpg,.jpeg} % for pdflatex we expect .pdf, .png, or .jpg files
\else%                                 % else we use pure latex
  \usepackage{graphicx}                % allow us to embed graphics files
  \DeclareGraphicsExtensions{.eps}     % for pure latex we expect eps files
\fi%

\graphicspath{{figures/}{pictures/}{images/}{./}} % where to search for the images

\usepackage{microtype}                 % use micro-typography (slightly more compact, better to read)
\PassOptionsToPackage{warn}{textcomp}  % to address font issues with \textrightarrow
\usepackage{textcomp}                  % use better special symbols
\usepackage{mathptmx}                  % use matching math font
\usepackage{times}                     % we use Times as the main font
\usepackage{cite}                      % needed to automatically sort the references
\usepackage{tabu}                      % only used for the table example
\usepackage{booktabs}                  % only used for the table example
\newcommand{\numVideo}{233}

\newcommand{\DD}{Data Level}
\newcommand{\DDI}{Image Level}
\newcommand{\DDO}{Object Level}
\newcommand{\DDE}{Event Level}
\newcommand{\DDC}{Tactic Level}

\newcommand{\DN}{Narrative Order}
\newcommand{\DNL}{Linear}
\newcommand{\DNFF}{FlashForward}
\newcommand{\DNFB}{FlashBack}
\newcommand{\DNZZ}{ZigZag}
\newcommand{\DNG}{Grouped}
\newcommand{\DNTF}{TimeFork}

\newcommand{\DV}{Visual Type}
\newcommand{\DVG}{Graphical Marks}
\newcommand{\DVVE}{Video Effects}

\newcommand{\DT}{Data Type}
\newcommand{\DTT}{Tracking Data}
\newcommand{\DTNT}{Non-tracking Data}

\newcommand{\EDU}{education}
\newcommand{\ENG}{entertainment}
\newcommand{\system}{VisCommentator}

\newcommand{\cmo}{\textcolor[rgb]{0, 0, 0}}
\newcommand{\rom}[1]{\uppercase\expandafter{\romannumeral #1\relax}}

\newcommand{\para}[1]{\vspace{1mm}\noindent\textbf{#1}}
\usepackage{enumitem}
\setlist[itemize]{noitemsep, topsep=0pt}
\usepackage{wrapfig}
\usepackage{caption}
\usepackage{etoolbox}
\usepackage{dblfloatfix}

\onlineid{1578}

\vgtccategory{Research}
\vgtcpapertype{Applications}

\title{Augmenting Sports Videos with \system{}}

\author{Chen Zhu-Tian, Shuainan Ye, Xiangtong Chu, Haijun Xia, Hui Zhang, Huamin Qu, and Yingcai Wu}
\authorfooter{
\item
 C. Zhu-Tian was with the State Key Lab of CAD \& CG, Zhejiang University and Hong Kong University of Science and Technology. He is now with the Department of Cognitive Science and Design Lab, University of California, San Diego. E-mail: zhutian@ucsd.edu.
\item
 S. Ye, X. Chu, and Y. Wu. are with the State Key Lab of CAD \& CG, Zhejiang University. E-mail: \{sn\_ye, chuxiangtong, ycwu\}@zju.edu.cn. Y. Wu is the corresponding author.
\item
 H. Xia is with the Department of Cognitive Science and Design Lab, University of California, San Diego. E-mail: haijunxia@ucsd.edu.
 \item
 H. Zhang is with the Department of Sport Science, Zhejiang University. E-mail: zhang\_hui@zju.edu.cn
\item
 H. Qu is with Hong Kong University of Science and Technology. E-mail: huamin@cse.ust.hk.
}

\shortauthortitle{Biv \MakeLowercase{\textit{et al.}}: Global Illumination for Fun and Profit}
\abstract{%% v0.6
Visualizing data in sports videos is gaining traction in sports analytics,
given its ability to communicate insights and explicate player strategies engagingly. 
However, augmenting sports videos with such data visualizations is challenging, 
especially for sports analysts,
as it requires considerable expertise in video editing.
To ease the creation process, 
we present a design space that characterizes augmented sports videos at an element-level (\emph{what the constituents are}) and clip-level (\emph{how those constituents are organized}). We do so 
by systematically reviewing \numVideo{} examples of augmented sports videos collected from TV channels, teams, and leagues. 
The design space guides selection of data insights and visualizations for various purposes.
Informed by the design space and close collaboration with domain experts, 
we design \system{}, a fast prototyping tool,
to eases the creation of augmented table tennis videos by
leveraging machine learning-based data extractors and design space-based visualization recommendations.
With \system{}, 
sports analysts can create an augmented video 
by \emph{selecting the data} to visualize 
instead of manually \emph{drawing the graphical marks}.
Our system can be generalized to other racket sports (\emph{e.g.}, tennis, badminton) once the underlying datasets and models are available.
A user study with seven domain experts shows high satisfaction with our system,
confirms that the participants can reproduce augmented sports videos in a short period,
and provides insightful implications into future improvements and opportunities.
}

\CCScatlist{ % not used in journal version
 \CCScat{K.6.1}{Management of Computing and Information Systems}%
{Project and People Management}{Life Cycle};
 \CCScat{K.7.m}{The Computing Profession}{Miscellaneous}{Ethics}
}

\teaser{
  \centering
  \includegraphics[width=0.95\textwidth]{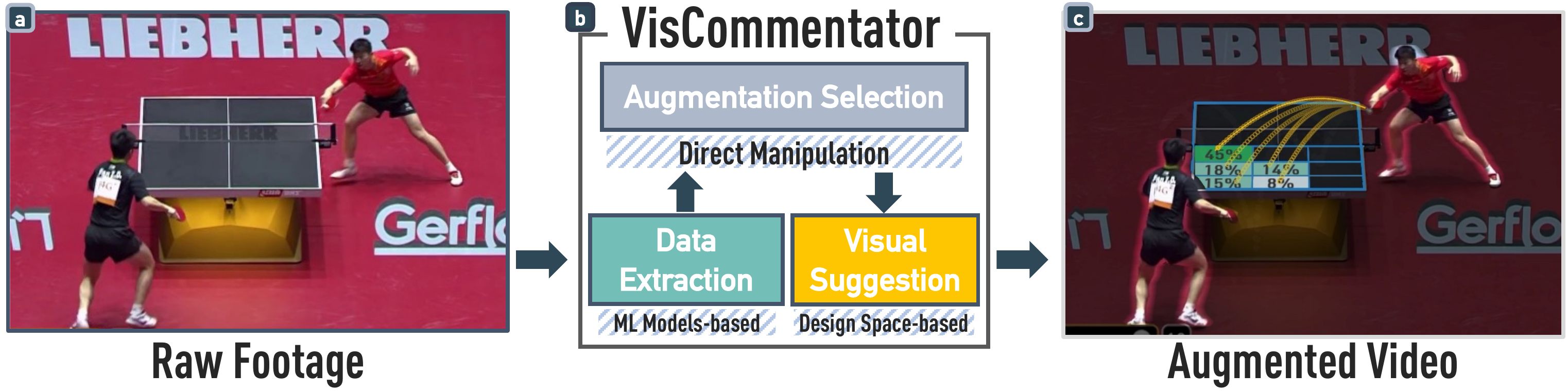}
    \caption{
    Left: A raw sports video.
    Middle: \system{} extracts data from the raw video by using machine learning models, allows users to select the augmenting data by directly interacting with the objects in the video,
    and suggests visuals for the data to generate an augmented video.
    Right: An augmented video reveals the probability distribution of the ball placement on the table.
    }
  \label{fig:teaser}
}

\vgtcinsertpkg

\begin{document}

\firstsection{Introduction}
\maketitle

%% why, who, when need augmented videos
Video is a popular medium for presenting and broadcasting sports events.
In recent years,
thanks to the advance of techniques such as Computer Vision (CV) and Image Processing (IP),
there has been a growing practice of augmenting sports videos with embedded visualizations~\cite{fischer2019video}.
These augmented sports videos combine visualizations
with video effects (\emph{e.g.}, slow motion, camera rotation)
to embed the data in the actual scenes (\autoref{fig:teaser}c).
Given the ability to explicate player strategies in an engaging manner,
augmented sports videos have been used widely by TV channels~\cite{espn} to engage the fans
and by sports analysts~\cite{perin2018state, stein2017bring} to communicate analytical findings.

Despite their popularity,
these augmented videos, by nature, are difficult to create,
especially for sports analysts, who
usually lack sufficient video editing skills compared to the proficient video creators at TV companies.
Based on our collaboration with sports analysts
who provide data analysis services for national sports teams,
we are informed that augmenting videos are particularly useful for sports analysts,
allowing them to effectively communicate the analytical insights to the coaching team and players.
However, sports analysts usually cannot create these videos
as the creation process involves complex design decisions and video editing.
Thus, fast prototyping tools that allow sports analysts to create augmented videos rapidly
are in great demand.

%% SOTA
Various systems and techniques have been proposed 
to augment sports video over the last years.
On the one hand,
industrial companies have developed commercial tools, 
such as Viz Libero~\cite{vizrt} and Piero~\cite{piero}, for TV sportscasts.
Yet, these tools target proficient video editors
and focus on the editing of graphical elements.
% and have a steep learning curve.
On the other hand,
although increasing research
has been conducted on sports data visualization,
visualizing data in sports videos has received relatively less attention from the community~\cite{perin2018state}.
A notable exception comes from Stein et al.~\cite{stein2017bring, stein2018},
who developed a system that 
automatically extracts and visualizes data from and in \emph{soccer} videos.
Nevertheless, except for soccer, there are no publicly available methodologies and tools
that help sports analysts create augmented sports videos 
by providing data-driven features (\emph{e.g.}, mapping and binding between data and visuals, visualizations recommendation).

% can be divided into two stages, namely, data analysis and visualization.
To facilitate the creation of augmented sports videos,
we are particularly interested in four questions that should be considered at two levels.
The first level, namely, \emph{element-level}, 
is about identifying the building blocks for augmenting sports videos.
Similar to all data visualizations,
the building blocks of augmented sports videos include both the data and visuals:
\emph{what kinds of data can be used to augment a sports video (Q1)?};
\emph{what visuals are used to present these data (Q2)?}
% After figuring out these constituents, 
The next level, \emph{clip-level}, is about the organization of these building blocks:
\emph{how to organize and present the data for various narrative purposes (Q3)?};
\emph{how to organize the visuals in temporal order with respect to the raw video (Q4)?}
Without regard for these considerations,
the visualizations embedded in the fast-moving video will easily overwhelm the audience, let alone engage them.

In this work, 
we focus on ball sports and aim to facilitate the creation of augmented videos for sports analysts.
To understand the design practices of augmented sports videos,
we first systematically reviewed a corpus of \numVideo{} examples
collected from reputable sources, such as TV channels, leagues, and teams. 
Our analysis resulted in a design space that characterizes augmented sports videos at element- and clip-levels,
which answer the aforementioned four questions.
The four design dimensions (\emph{i.e.}, \DT{}, \DV{}, \DD{}, and \DN{}) of the design space
and their combination frequency
provide guidance for selecting data insights and visual effects for various purposes, such as entertainment or education.
Informed by the design space, 
we design and implement \system{},
a proof-of-concept system 
that allows sports experts to rapidly prototype augmented table tennis videos.
\system{} takes a raw table tennis video as the input,
extracts the data based on data levels by using machine learning (ML) models,
allows users to select the augmenting data 
by directly interacting with the objects in the video,
and suggests visual effects 
based on the narrative order and data selected by the user (\autoref{fig:teaser}b).
% Integrated with existing video editors,
% \system{}
Due to the data-driven nature, \system{} can be generalized to other racket ball sports 
such as tennis and badminton once data and the underlying models are available.
%% evaluation
A reproduction-based user study with seven sports analysts
demonstrates the overall usability of the system.
We further report the analysts’ feedback gathered from the post-study interviews,
which implies future improvements and opportunities.
In summary, the main contributions of this research include 
1) a design space derived from existing augmented sports videos,
2) the design and implementation of a fast prototyping tool for augmenting table tennis videos, 
and 3) a reproduction-based user study with seven domain experts.
The corpus, created videos, and other materials can be found in \url{https://viscommentator.github.io}.
\section{Related Work}

\vspace{-1mm}
\para{Video-based Sports Visualization.}
Due to its advantages of presenting data in actual scenes,
video-based sports visualization has been used widely to
ease experts' analysis~\cite{stein2017bring} 
and engage the audiences~\cite{lin2020sportsxr}.
Based on the presentation method, 
video-based visualizations
can be divided into side-by-side~\cite{Zhi2019},
overlaid~\cite{tang2020design},
and embedded~\cite{stein2017bring}.
This work focuses on embedded visualizations in sports videos.

Perin et al.~\cite{perin2018state} 
comprehensively surveyed the visualization research on sports data,
indicating that only a few works can be considered as video-based visualizations.
Among these few works,
a representative example was developed by Stein et al.~\cite{stein2017bring} for soccer.
Their system takes raw footage as the input 
and automatically visualizes certain tactical information as graphical marks in the video.
Later, Stein et al.~\cite{stein2018} extended their work 
by proposing a conceptual framework that semi-automatically
selects proper information to be presented at a given moment.
Recently, 
Fischer et al.~\cite{fischer2019video} found that video-based soccer visualization systems from the industry are actually ahead of most of the academic research.
For example, 
Piero~\cite{piero} and Viz Libero~\cite{vizrt} are both developed for sportscasting
and provide a set of powerful functions to edit and annotate a sports video.
Besides, CourtVision~\cite{courtvision}, developed by Second Spectrum~\cite{secondspectrum} for basketball, 
is another industry product that automatically tracks players' positions 
and embeds status information to engage the audiences.
In summary, as detailed by Fischer et al.~\cite{fischer2019video},
the research on video-based sports visualization is still in its infancy, 
while the strong market demand has already spawned very successful commercial systems.
Nevertheless, 
these commercial systems target proficient video editors,
leading to a steep learning curve for sports analysts.
Moreover, they %rarely enable data-driven designs whereas
mainly augment the video with graphical elements,
while the goal of sports analysts is to augment sports videos with data.
Given that very little is known about the design practices of augmented sports videos, 
it remains unclear how to support augmenting sports videos with data.
In this work, we explore this direction and aim to ease the creation
by a systematic study of existing augmented sports videos collected
from reputable sources.

\vspace{1mm}
\para{Data Videos for Storytelling.}
Data video, 
as one of the seven genres of narrative visualization categorized by
Segel and Heer~\cite{narrativeVis},
is an active research topic and has attracted interest from researchers.
Amini et al.~\cite{amini2015} systematically analyzed 50 data videos from reputable sources and summarized the most common visual elements and attention cues in these videos.
They further deconstructed each video into four narrative categories
by using the visual narrative structure theory introduced by Cohn~\cite{cohn2013}.
Their findings reveal several design patterns in data videos 
and provide implications for the design of data video authoring tools.
Build upon that, Amini et al.~\cite{Amini2017} further contributed DataClips,
an authoring tool that allows general users to craft data videos with predefined templates.
Recently, 
Thompson et al.~\cite{Thompson2020} contributed a design space of data videos
for developing future animated data graphic authoring tools.
Cao et al.~\cite{cao2020} analyzed 70 data videos and proposed a taxonomy to
characterize narrative constructs in data videos.

%% Embedded
Although these studies provide insights into data video design,
our scenario is inherently different from theirs and thus yields new challenges.
Specifically,
these studies use video as a medium to tell the story of data,
whereas we focus on augmenting existing videos with data.
An existing video imposes extra constraints 
on narrative orders, visual data forms, \emph{etc.}
With these constraints,
how to visually narrate data in videos remains underexplored.
Perhaps the most relevant work is from Stein et al.~\cite{stein2018},
who annotated the data in soccer videos in a linear way.
To fully explore the ways for augmenting sports videos with data,
we analyze \numVideo{} real-world examples 
and summarize six narrative orders and their common usage scenarios.
The results provide guidance for designing 
authoring tools for augmented sports videos.

% \vspace{1mm}
\para{Intelligent Design Tools.}
Designing visual data stories usually requires the skills to analyze complex data 
and map it into proper visualizations,
both of which require considerable skills.
Therefore, to lower the entry barrier to visual data stories,
many researchers develop intelligent creation tools 
to automate or semi-automate the design process.
One widely used approach to ease the visual mapping process is templates. 
For example,
DataClips~\cite{Amini2017} and Timeline StoryTeller~\cite{brehmer2017} use templates manually summarized from existing examples to enable semi-automatic creation of data animation and timeline infographics, respectively.
On the basis of Timeline StoryTeller, Zhu-Tian et al.~\cite{chentimeline} proposed a method to automatically extract extensible templates from existing designs
to automate the timeline infographics design.
Besides automating the visual mapping process,
some tools further facilitate the data analysis process 
by automatically suggesting data insights.
%% Data-shot
DataShot~\cite{WangSZCXMZ20} adopts an auto-insight technique to recommend interesting data insights based on their significance for factsheet generation.
%% Data-toon
DataToon~\cite{kim2020}, an authoring tool for data comic creation, 
uses a pattern detection engine 
to suggest salient patterns of the input network data.

However, few, if any, tools exist that provide the aforementioned kinds of
data-driven support for creating augmented sports videos.
The challenges of developing such a tool not only exist 
in the engineering implementations but also in the integration
between the workflows of visualization authoring and video editing.
We draw on the line with prior visualization design tools
and design \system{} to support visualizing data sports videos in a video editing process.
% in a video editing.

% \vspace{1mm}
\para{Data Extraction from Sports Videos.}
Due to the advancement of deep learning on CV,
such as object detection~\cite{Ren2015}, segmentation~\cite{Redmon2015}, 
and action recognition~\cite{zhang2019pan}, 
more and more data is available to be extracted from a video~\cite{EmoCo, EmoCue, aoyu, haotian}.
Shih~\cite{shih2017survey} 
presented a comprehensive survey on content-aware video analysis for sports.
He summarized the state-of-the-art techniques to parse sports videos
from a low-semantic to a high-semantic level:
1) object level techniques extract the key objects in the video, 
2) event level recognizes the action of the key object, 
and 3) conclusion level generates the semantic summarization of the video.
Besides models tailored to sports videos,
researchers have also developed models~\cite{nguyen2016} to detect and segment humans from general images and videos.
A particular challenge in table tennis is that
the ball is small (15 pixels on average~\cite{ttnet}) 
and move fast (\emph{e.g.}, $30$ $m/s$).
To tackle this challenge, Voeikov et al.~\cite{ttnet}
proposed TTNet, a convolution-based neural network,
to detects the ball in a high-resolution video.
Besides, TTNet can further segments the ball 
and detect ball events such as bounces and net hits.
In this work,
we composite multiple state-of-the-art machine learning models~\cite{ttnet, bodypiex2, he2016}
to extract the data from table tennis videos, % with a fixed camera angle.
such as ball and players' position, actions, events, and key strokes.

\section{Design space of augmented videos}
To understand the design practices of augmented videos for ball sports,
we collected and analyzed \numVideo{} videos from TV channels, teams, and leagues
to characterize augmentations 
at element-level (\emph{what the constituents are}) 
and clip-level (\emph{how those constituents are organized}).

\subsection{Methodology}

\begin{figure*}[th]
  \centering
  \includegraphics[width=\textwidth]{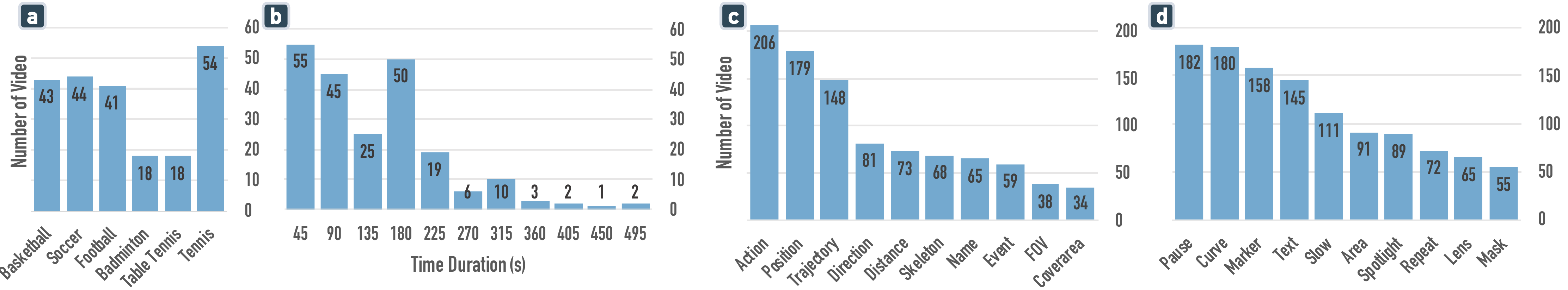}
  \vspace{-2mm}
  \caption{
    The number of videos a) of different ball sports and b) with different time duration.
    The top-10 frequently presented c) data types and d) visual types in augmented sports videos.
  }
  \vspace{-5mm}
 \label{fig:count}
\end{figure*}

\begin{figure*}[b]
  \centering
  \vspace{-2mm}
  \includegraphics[width=\textwidth]{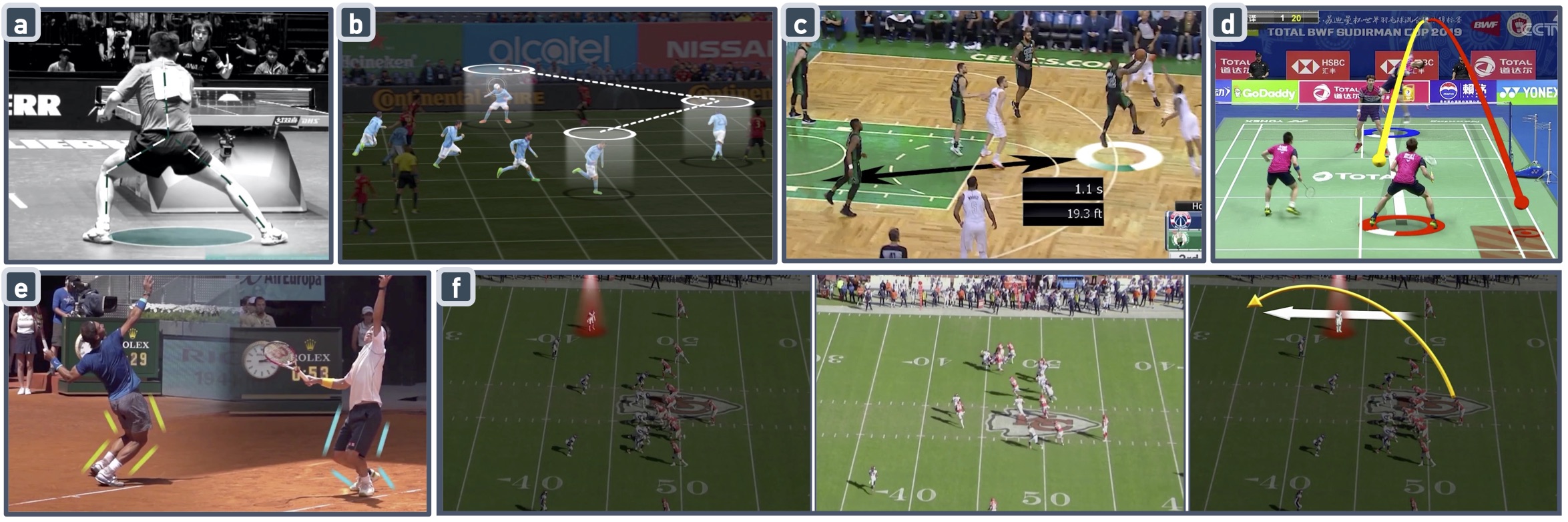}
  \caption{Video examples in the corpus of a) table tennis in Linear, b) soccer in FlashForward, c) basketball in FlashBack, d) badminton in TimeFork, e) tennis in Grouped, and f) football in ZigZag. }
 \label{fig:examples}
\end{figure*}

%% the data source and statistics
% \paragraph{\textbf{Data Sources and Videos.}} 
\vspace{-1mm}
\para{Data Sources and Videos.}
We collected a video corpus of six popular ball sports,
including three team sports (\emph{i.e.}, basketball, soccer, and American football)
and three racket sports (\emph{i.e.}, tennis, badminton, and table tennis).
 Specifically, we searched the videos in Google Videos
by using keyword combinations such as ``breakdown videos + SPORT'', ``analysis video + SPORT'', and ``AR + SPORT'',
where SPORT is one of the six ball sports mentioned above.
The searching processing was repeated recursively on each returned site 
until no more augmented videos were found.
To ensure the quality and representative of the videos, 
we only included videos that are 
watched by more than tens of thousands of times
and created by official organizations, such as TV companies, sports teams.
We also purchased some subscriptions (such as ESPN+, CCTV VIP)
to watch member-only videos during the collection.
Three of the authors went through the videos to exclude the problematic ones
(\emph{e.g.}, with no or only a few augmentations, not ball sports, or not cover a sports event).
In this process, we noticed that most of the videos focused on one sports event (\emph{e.g.}, a goal) 
and thus were less than 3mins.
Some videos were sports events collections and too long (\emph{e.g.}, 45mins).
Thus, we sliced these videos into pieces to ensure that each piece is shorter than 3mins 
and includes at least one sports event.
To control the diversity of the videos,
we randomly sampled a subset of videos from the raw corpus (which contains more than $1000$ videos)
following this priority: 
the number of 1) team sports \emph{vs.} racket sports, 
2) different sports types, 
and 3) different video sources.
Finally, our corpus includes \numVideo{} videos. 
\autoref{fig:count}a and b present 
the number of videos of different sports types
and with different time duration.
\autoref{fig:examples} presents some examples from our corpus.

\para{Qualitative Analysis.}
% We are particularly interested in the augmentations
% of the collected videos from 1) element-level (\emph{i.e.}, \emph{what are the constituents of the augmentations?}) 
% and 2) clip-level (\emph{i.e.}, \emph{how the constituents are organized?})
% To this end, 
We conducted a qualitative analysis of this corpus.
Specifically, we first followed the definition in~\cite{amini2015} to segment the augmented sports videos into clips.
In our corpus, a video can have more than one sports event,
each of which can be segmented into multiple clips.
Three of the authors independently reviewed and segmented
the videos following this process:
each of them segmented $1/3$ videos
and verified another $1/3$ videos segmented by others.
Gathering the reviews resulted in 871 clips.
We then probed into the content of 261 (30\%) clips \cmo{to examine} the element-level design.
The investigation also referred to prior research~\cite{perin2018state, amini2015, Thompson2020}
and derived the code-sets of two element-level dimensions, \DT{} and \DV{}.
Next, three of the authors independently conducted open coding on the 261 clips
to characterize how these visual elements are organized.
Disagreements were resolved through multiple rounds of discussion
and iterations with another three authors,
one of whom was a senior sports analyst and had more than ten years' experience
in providing consulting services for national sports teams and sports TV channels,
until we reached a Cohen's $\kappa$ above 0.7 for all codes of videos.
Finally, we came to two clip-level dimensions, namely, \DD{} and \DN{}.
The design dimensions were further refined in the following process.

\para{Annotations.}
We further counted the occurrences of these design dimensions
in the corpus.
%%%%%%%%% annotations
Five postgraduate students were recruited to annotate these videos following our codes.
All of them were sports fans.
We introduced the details of the design dimensions to the five students,
asked them to practice annotation on 15 curated example videos,
and started the actual annotation when they were confident enough.
After the annotation, cross-validation was further conducted.
In total, each student annotated 45 videos and validated 45 videos from two others.
Questions and discussions were encouraged throughout the process, 
which helped us refine the design dimensions.
Finally, three of the authors scanned through the annotations and calculated the statistics.
% The statistics are summarized in Sec.\ref{ssec:design_patterns}.
\setlength{\intextsep}{0pt}
\setlength{\columnsep}{3pt}

\subsection{Element-level Design}
\label{ssec:design_element}

% Augmented sports videos are augmented by visual elements.
% The augmentations of a augmented sports video are achieved by visual elements.
The augmentations of a sports video are presented as visual elements.
We characterize these visual elements with \DT{} and \DV{}.

\para{Dimension \rom{1}: \DT{}.}
% As a kind of sports visualization,
% a fundamental characteristics of augmented sports videos 
% is to augment raw sports videos with data
% and ultimately communicate insights with or engage the audiences.
% Behind the visual elements is data.
% The visual elements are driven by data.
Sports videos are augmented with different types of data.
Perin et al.~\cite{perin2018state} 
identified three types of data used in sports visualizations: 
tracking (in-game actions and movements), 
box-score (historical statistics),
and meta (sports rules and player information).
Given the in-game nature of sports videos,
most of the augmented data on a sports video belongs to tracking data.
Thus, we simplify the three types into two:

\begin{itemize}[leftmargin=*]
    \item \emph{\underline{\DTT{}}} is the data collected or extracted from a specific game, such as the moving trajectories and the actions of players. 
    In recent years, the advances of CV techniques lead to increased tracking data that ranges from low-level physical data to high-level tactical data. 
    This data is always associated with a specific space and time in the video and can be naturally embedded into the video. Thus, most of the data visualized in augmented videos is tracking data. 
    
    \item \emph{\underline{\DTNT{}}} refers to the data not captured from a specific game, including historical data, rules, and player information. 
    Augmented videos usually provide this data as supplemental information to explain or comment on the situations in the game.
\end{itemize}

\noindent
\autoref{fig:count}c shows the top-10 frequently presented data types in our corpus.
Most of (9/10) the data is tracking data except for \emph{name}, which belongs to non-tracking data.

\para{Dimension \rom{2}: \DV{}.}
The data is presented as different types of visuals.
Usually, in data videos, data is presented as graphical marks~\cite{Amini2017},
such as bar, pictograph, and map.
However, in augmented sports videos,
the video content in the raw footage such as players and the court
can also be used to encode data.
We categorize these visual representations as \DVG{} and \DVVE{}:

\begin{itemize}[leftmargin=*]
    \item \emph{\underline{\DVG{}}} are the visual elements added to the raw footage.
    Augmented sports videos usually present data as primitive marks, 
    such as dots, lines, and areas, while common data videos~\cite{amini2015} 
    contain more complex visualizations, such as donut, pie, scatter plot.
    We also see some dedicated marks that rarely have been used in common data videos, such as \emph{spotlight}, \emph{skeleton}, and \emph{field of view}. 
    Besides, we only found two types of animation used for the marks in augmented sports videos, namely, \emph{Creation} and \emph{Destruction}, while eight~\cite{Amini2017} are used in common data videos.

    \item \emph{\underline{\DVVE{}}} are visual effects based on the existing content of the raw footage, such as segmenting and moving a player, showing a slow-motion, or rotating the camera.
    In a certain aspect, these effects are similar to the Attention Cues in~\cite{amini2015}.
    However, in augmented sports videos, these effects are driven or controlled by data,
    with the goal of not only drawing the viewer's attention but also revealing deeper
    insights into a sports event.
    For example, the video can move a player to show a what-if situation
    or rotate the camera to present another point-of-view.
    These effects are limited by the quality of the raw footage (frame rate, resolution) and video processing techniques.

\end{itemize}

\noindent
\autoref{fig:count}d shows the top-10 frequently used visual types in our corpus.
7/10 are graphical marks, while only 3 belong to video effects (\emph{pause}, \emph{slow}, and \emph{repeat}).

\subsection{Clip-level Design}
\label{ssec:design_clip}

\begin{figure*}[ht]
  \centering
  \includegraphics[width=0.75\textwidth]{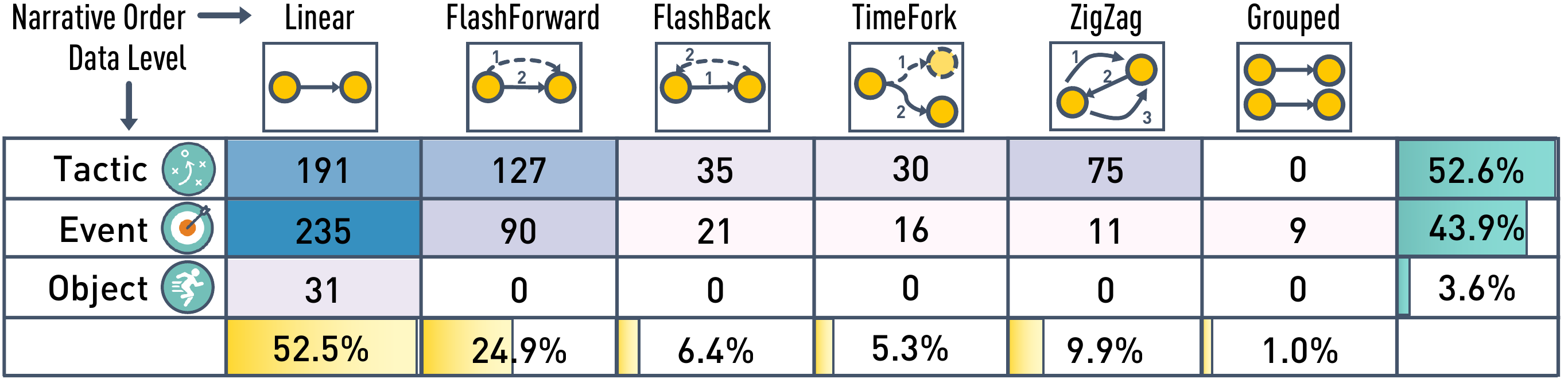}
  \vspace{-2mm}
  \caption{
    A clip-level design space for augmented sports videos:
    Data Level and Narrative Order. 
    The number in cells depict their combination occurrences in our corpus. 
    Darker cells mean more occurrences. 
    The last row and column present the ratio of each option to its dimension.
  }
 \label{fig:pattern_table}
 \vspace{-5mm}
\end{figure*}

A more important question is how to organize the data and visual types in a sports video.
To cope with this question,
we identify two design dimensions, \DD{} and \DN{},
that should be considered for selecting and presenting data in the video.

\para{Dimension \rom{3}: \DD{}.}
Augmented videos usually present sports data for different purposes.
Specifically, some videos present data for 
\emph{\ENG{}} (\emph{e.g.}, \emph{showing the jumping height of a player who is dunking})
while others are for 
\emph{\EDU{}} (\emph{e.g.}, \emph{highlighting the formation of the team}).
We notice that the data for 
\ENG{} is usually with low semantics (\emph{e.g.}, positions and distances)
that can be easily perceived from the raw footage by general audiences,
while the one for 
\EDU{} is more often with high semantics (\emph{e.g.}, techniques and tactics) 
that provide extra knowledge to the audiences.
Based on this observation, we categorize the data from low to high semantic levels:

\begin{wrapfigure}[2]{l}{0.04\textwidth}
  \begin{center}
    \vspace{-12pt}
    \includegraphics[width=0.04\textwidth]{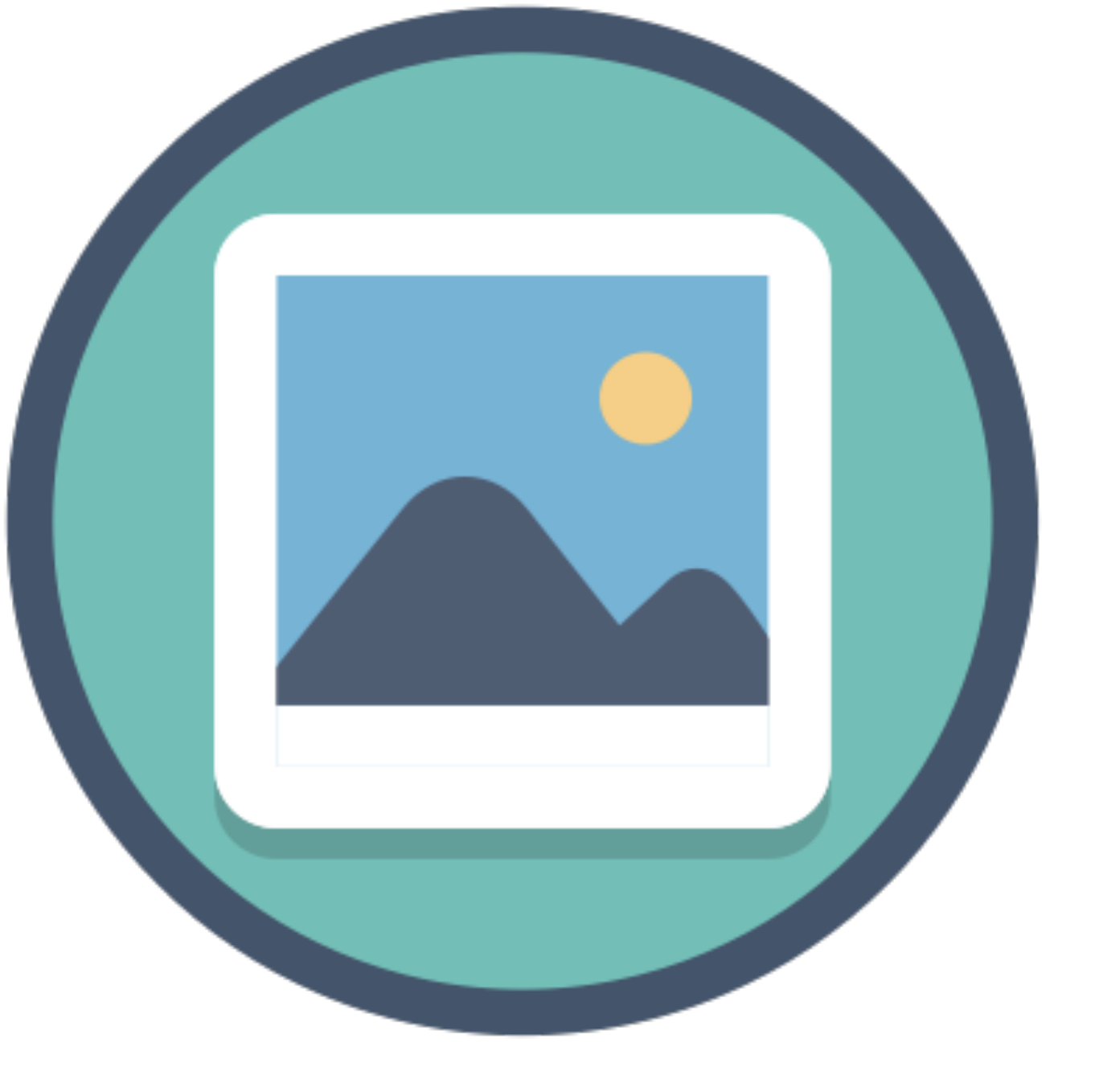}
  \end{center}
\end{wrapfigure} 
\noindent
\emph{\underline{\DDI{}}} includes the frames of a raw footage. The data at this level has the largest quantity and will be used as the input of a system. A video without any augmentations presents this level's data. 

\begin{wrapfigure}[2]{l}{0.04\textwidth}
  \begin{center}
    \vspace{-12pt}
    \includegraphics[width=0.04\textwidth]{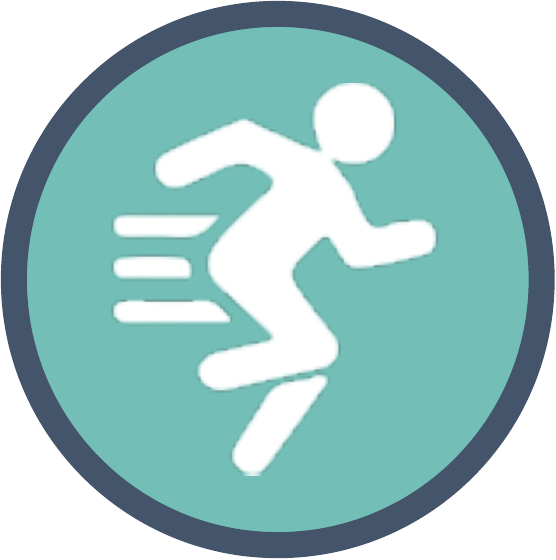}
  \end{center}
\end{wrapfigure} 
\noindent
\emph{\underline{\DDO{}}} includes the physical data of objects detected from the image-level data, such as the postures of a player, the positions of the ball, and empty areas of the court. This level of data can be naturally understood from the video by audiences without any prior knowledge. An augmented video that only presents the data at this level is mainly \textbf{for \ENG{} purposes}.

\begin{wrapfigure}[2]{l}{0.04\textwidth}
  \begin{center}
    \vspace{-12pt}
    \includegraphics[width=0.04\textwidth]{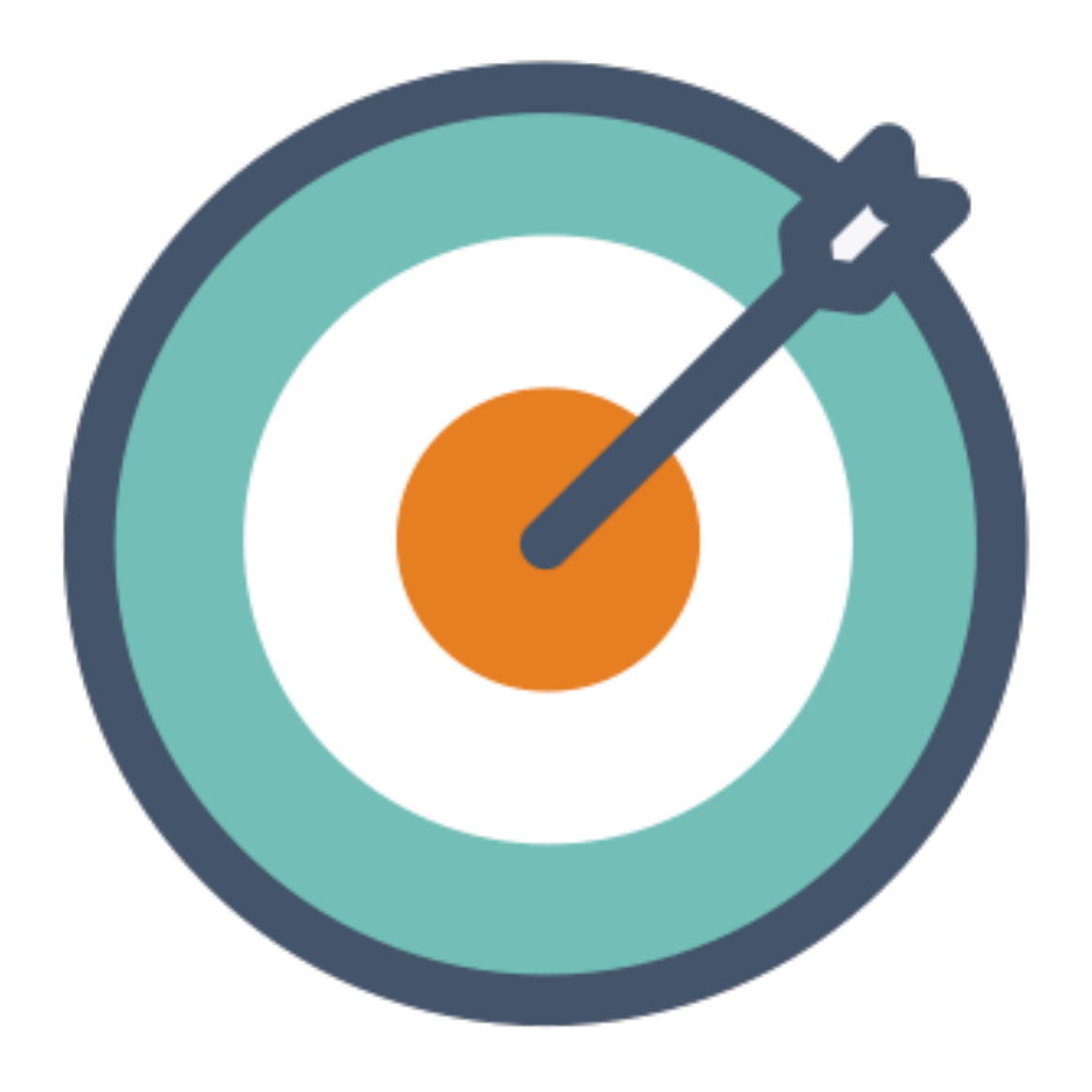}
  \end{center}
\end{wrapfigure} 
\noindent
\emph{\underline{\DDE{}}} contains the data that can only be interpreted from the video with domain specific knowledge. 
Typical examples include the player's techniques, the formation of the team, and the status of the ball. These kinds of data may be familiar to experienced fans but provide extra knowledge to novices. 
Therefore, when presenting event-level data, an augmented video is considered to be in the middle \textbf{between for \ENG{} and for \EDU{}}.

\begin{wrapfigure}[2]{l}{0.04\textwidth}
  \begin{center}
    \vspace{-12pt}
    \includegraphics[width=0.04\textwidth]{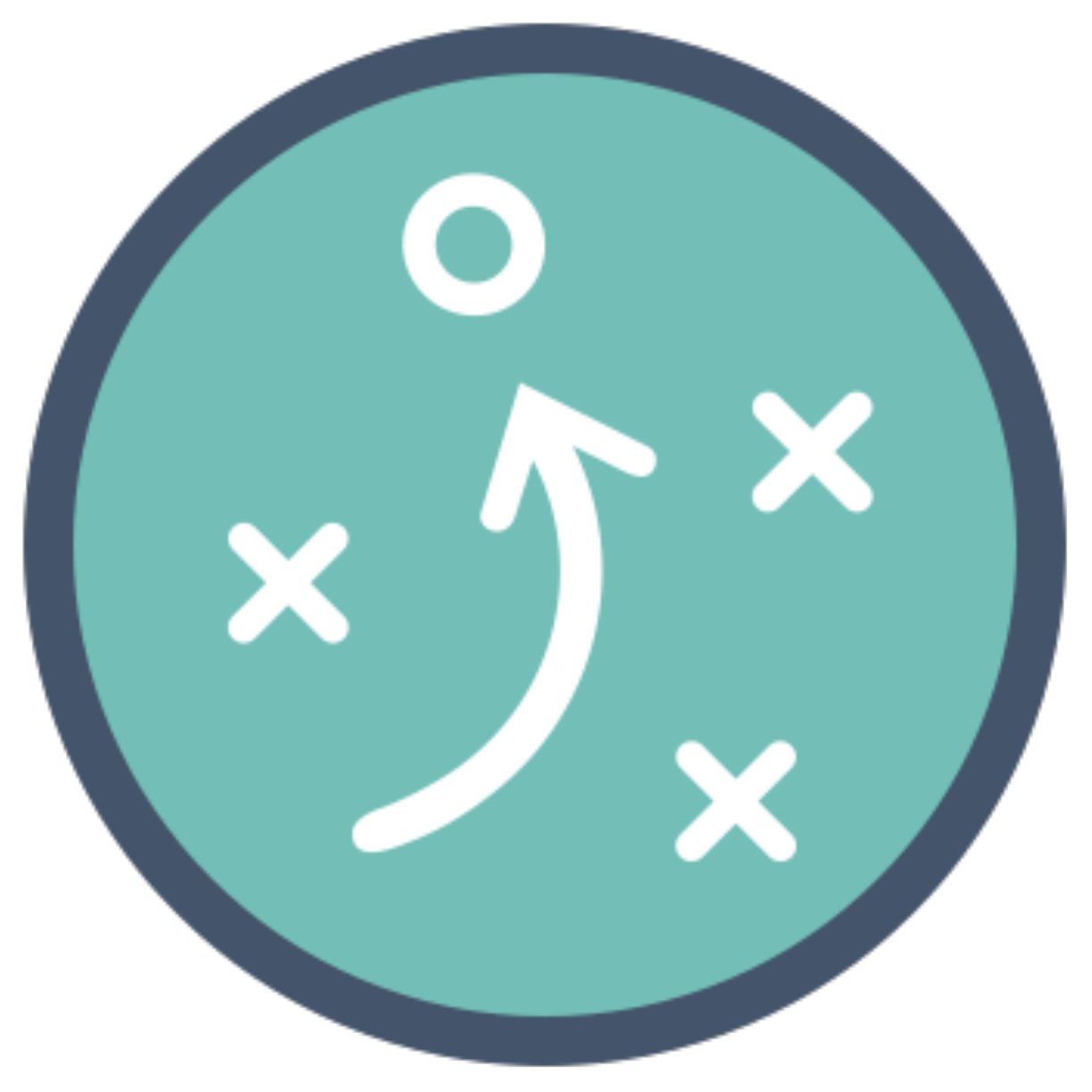}
  \end{center}
\end{wrapfigure} 
\noindent
\emph{\underline{\DDC{}}} presents the reasoning results of a sports video, explaining why a team wins or loses a score. 
% Usually an augmented video contains at most one conclusion (\emph{i.e.}, why one side wins or loses).
Tactic-level data is usually drawn by experts who analyze the data from other levels, indicating the key factors that lead to the result.
Hence, an augmented video with tactic-level data is mainly \textbf{for \EDU{} purposes} since it provides the most additional knowledge for audiences.

\autoref{fig:pattern_table} shows the statistics of clips with different data levels.
Overall, most of the clips present tactic- (52.6\%) and event-level (43.9\%) data,
while only 3.6\% of clips present object-level data. 
These statistics indicate that most the augmented videos 
target sports fans and provide expert analysis.
\DD{} is not meant to be a taxonomy of sports data
but a design dimension that needs to be considered in creating an augmented video.
For example, if a video designer wants to engage the novice audience, she should select and present the object-level data. 

\para{Dimension \rom{4}: \DN{}.}
A sports video usually presents the sports events in linear
and rarely employs non-linear narrative structures.
However, we notice that the data in these videos are not always presented in chronological order.
For example, to explain the tactic of a player,
it is common to pause the video and
foreshadow the trajectories of the next several movements of the player.
Considering the presenting order of data 
and its actual chronological order,
we borrow the idea of \emph{narrative order}~\cite{narrativeorder}
to depict how the data is presented in these augmented videos:

\begin{wrapfigure}[2]{l}{0.05\textwidth}
  \begin{center}
    \vspace{-12pt}
    \includegraphics[width=0.05\textwidth]{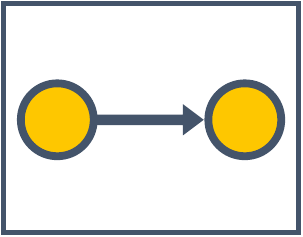}
  \end{center}
\end{wrapfigure} 
\noindent
\emph{\underline{\DNL{}}} presents data in chronological order (\autoref{fig:examples}a),
which is the most common way to present data in sports videos.
Pausing is used in \DNL{} when there is too much data to show in one moment.

\begin{wrapfigure}[2]{l}{0.05\textwidth}
  \begin{center}
   \vspace{-12pt}
    \includegraphics[width=0.05\textwidth]{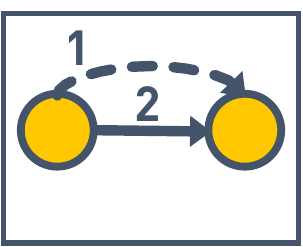}
\end{center}
\end{wrapfigure} 
\noindent
\emph{\underline{\DNFF{}}} foreshadows the data that will happen later than what is being told. 
    It is frequently used for \DDC{} data.
    The video is usually paused when using \DNFF{}. For example, % \emph{e.g.}, 
    \autoref{fig:examples}b shows the positions of one player
    in the next several seconds.

\begin{wrapfigure}[2]{l}{0.05\textwidth}
  \begin{center}
     \vspace{-12pt}
    \includegraphics[width=0.05\textwidth]{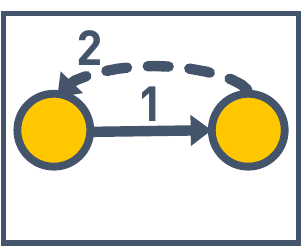}
  \end{center}
  \end{wrapfigure} 
 \noindent
     \emph{\underline{\DNFB{}}} presents the data that took place earlier than what is happening.
    \DNFB{} can be used for both \ENG{} (\emph{e.g.}, to emphasize the achievements of a player)
    and \EDU{} (\emph{e.g.}, explain the causality).
    For example, \autoref{fig:examples}c highlights a player's previous position to reveal how fast a player ran.
    
\begin{wrapfigure}[2]{l}{0.05\textwidth}
  \begin{center}
     \vspace{-12pt}
    \includegraphics[width=0.05\textwidth]{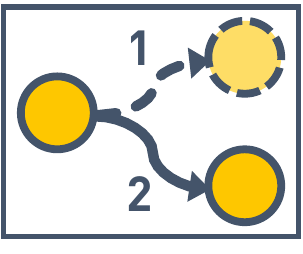}
  \end{center}
  \end{wrapfigure} 
 \noindent
    \emph{\underline{\DNTF{}}} shows data that never happens in the game 
    and is primary used to present a what-if analysis, 
    explaining the results of different choices of players (\autoref{fig:examples}d). 
    A typical pattern in \DNTF{} is to show the hypothetical data first,
    reject these options,
    and visualize the actual data at the end.

\begin{wrapfigure}[2]{l}{0.05\textwidth}
  \begin{center}
     \vspace{-12pt}
    \includegraphics[width=0.05\textwidth]{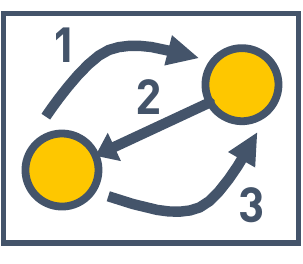}
  \end{center}
  \end{wrapfigure} \noindent
     \emph{\underline{\DNZZ{}}} plays the video in reverse for a short period 
    and then forwards this part again (\autoref{fig:examples}f).
    Basic usage of \DNZZ{} is to highlight some key events, 
    such as crossover and spin,
    for both \ENG{} and \EDU{} purposes.

\begin{wrapfigure}[2]{l}{0.05\textwidth}
  \begin{center}
     \vspace{-12pt}
    \includegraphics[width=0.05\textwidth]{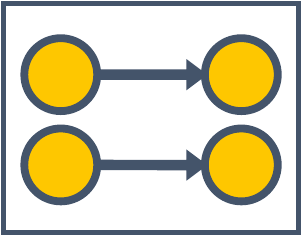}
  \end{center}
  \end{wrapfigure} 
 \noindent
     \emph{\underline{\DNG{}}} presents multiple sports scenes 
    by using picture-in-picture or multiple views (\autoref{fig:examples}e).
    These events are grouped based on some criteria, such as technique types, players, or correlation,
    and can be played in parallel or series.

\autoref{fig:pattern_table} provides the statistic of these \DN{} 
in the video corpus. 
In general, \DNL{} (52.5\%) and \DNFF{} (24.9\%) 
are the two most frequently used order,
Grouped (1\%) is the least frequently used one,
and the remaining three---\DNFB{}, \DNZZ{}, and \DNTF{}---all have around 5\% to 10\% cases.

\subsection{Patterns at Clip-level Design}
\label{ssec:common}

Based on the annotations of our corpus,
we identify some common design patterns at clip-level:

%% T/E/P + L
\begin{wrapfigure}[2]{l}{0.13\textwidth}
  \begin{center}
    \vspace{-12pt}
    \includegraphics[width=0.04\textwidth]{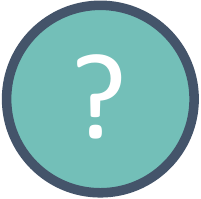}
    \includegraphics[width=0.03\textwidth]{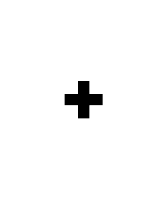}    
    \includegraphics[width=0.05\textwidth]{figs/dn_l.pdf}
  \end{center}
\end{wrapfigure} 
\noindent
 \emph{\underline{Data + \DNL{}.}} \emph{Linear} is the most commonly used (52.5\% clips) order to augment sports videos in all data levels.
 Presenting data in linear is natural since the underlying sports videos are played in linear. Thus, an augmented video creation tool can use the \DNL{} order as the default choice for all data levels.
 An interesting finding is that object-level data is seldomly narrated in orders other than \DNL{}.
 After discussing with our experts, we consider this is primary because:
 1) the physical data can easily be understood by the audiences,
 and 2) when augmenting a video with physical data to engage the audiences, 
 using linear without pausing can \textbf{avoid interfering the audiences' game-watching experience}.
 
 \begin{wrapfigure}[2]{l}{0.13\textwidth}
  \begin{center}
    \vspace{-12pt}
    \includegraphics[width=0.04\textwidth]{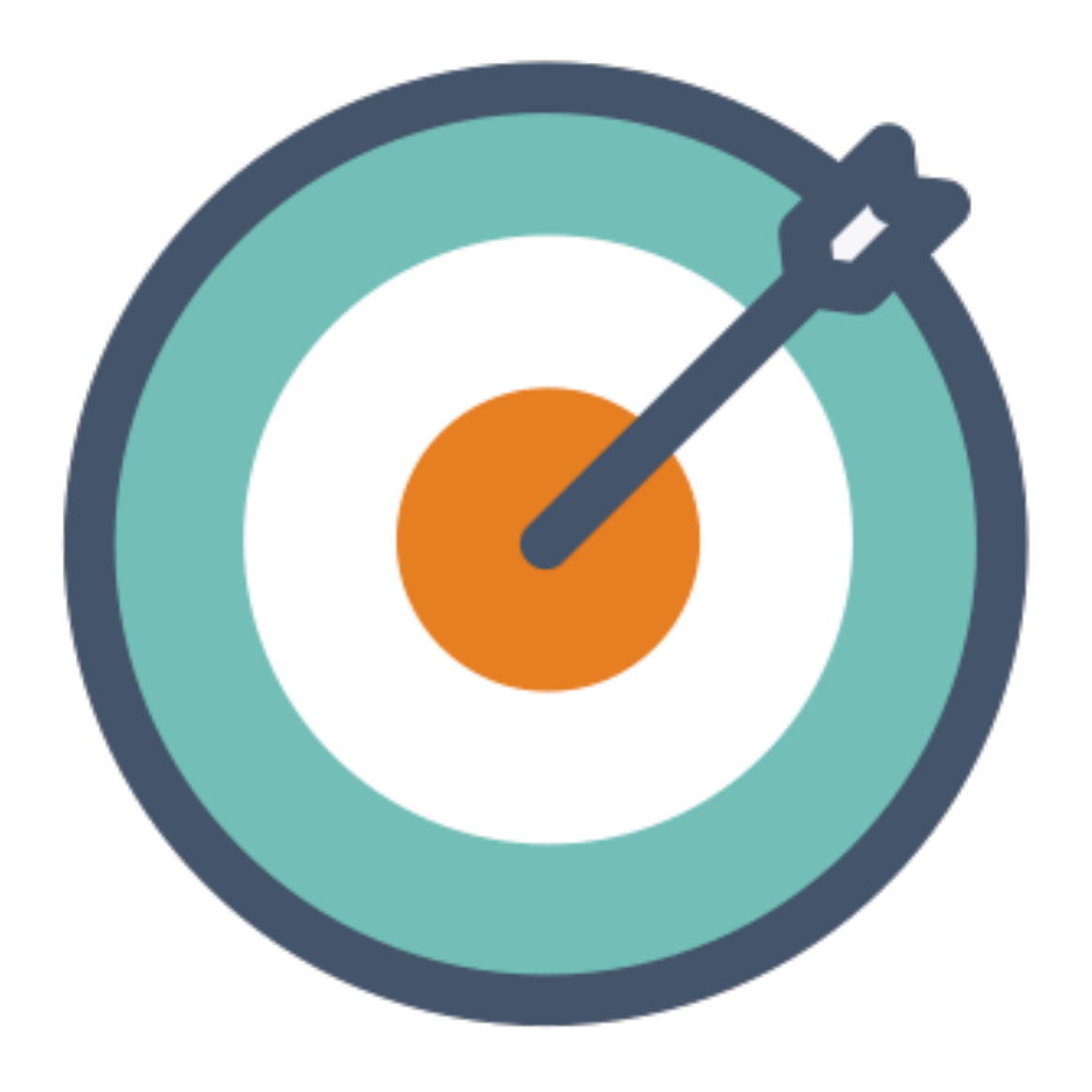}
    \includegraphics[width=0.03\textwidth]{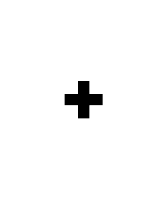}    
    \includegraphics[width=0.05\textwidth]{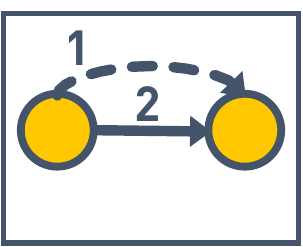}
  \end{center}
\end{wrapfigure} 
\noindent
\emph{\underline{\DDE{} + \DNFF{}.}} 
We also see that many clips show event-level data using \DNFF{}.
Through watching the videos in our corpus,
we notice that using \DNFF{} allows to 
\textbf{preview and explain complex data} in the following events 
% (\emph{e.g.}, in the play-by-play breakdown)
\textbf{without affecting the actual watching experience}. % when actually watching it}.

%% T + FF
\begin{wrapfigure}[2]{l}{0.13\textwidth}
  \begin{center}
    \vspace{-12pt}
    \includegraphics[width=0.04\textwidth]{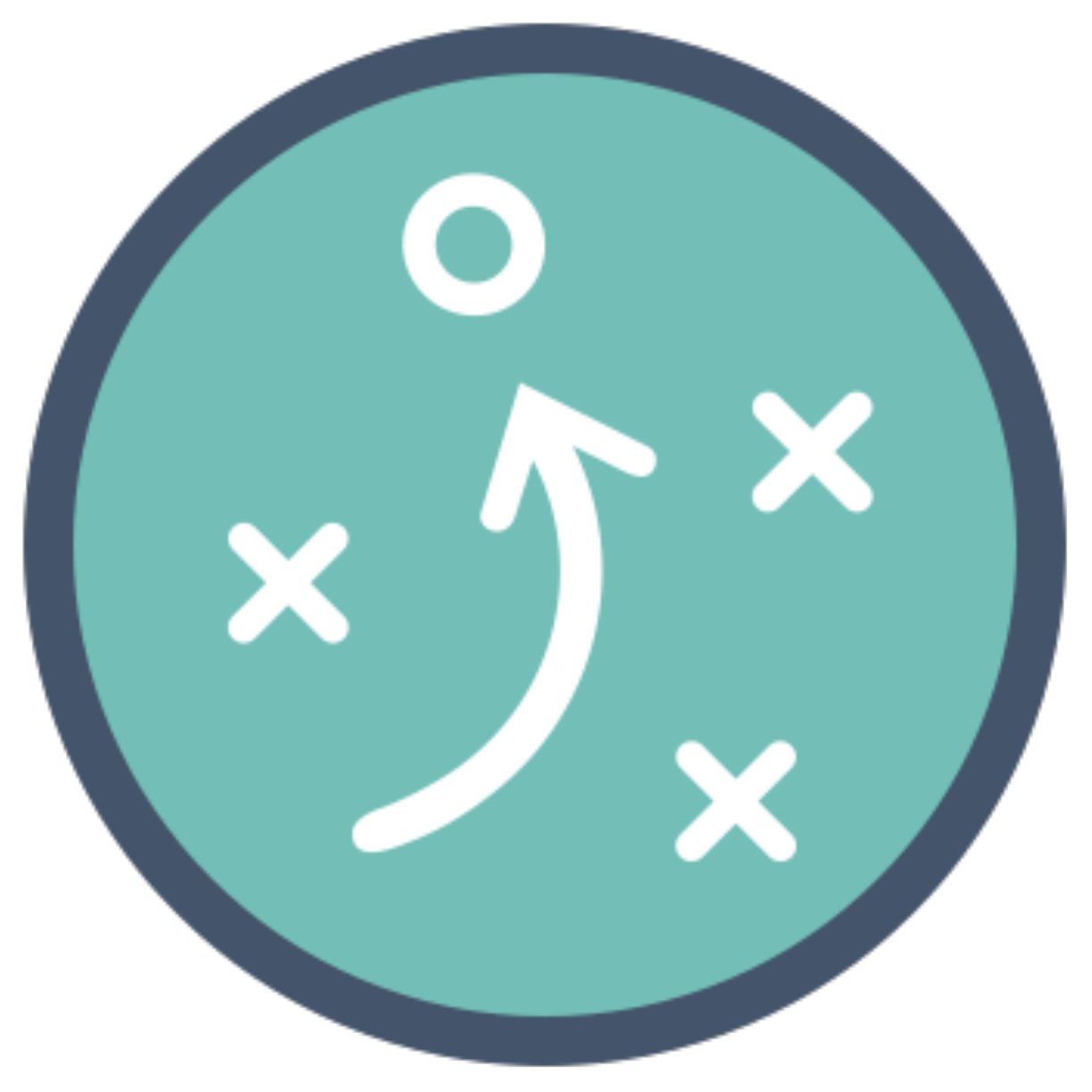}
    \includegraphics[width=0.03\textwidth]{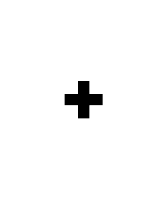}    
    \includegraphics[width=0.05\textwidth]{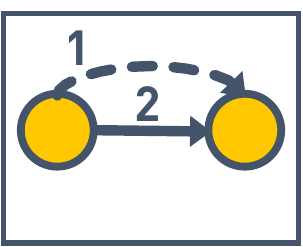}
  \end{center}
\end{wrapfigure} 
\noindent
\emph{\underline{\DDC{} + \DNFF{}.}} 
An interesting pattern is that compared to event-level data,
tactic-level data is less frequently presented in \DNL{} but more in \DNFF{}. 
By digging into the corpus, %%clips in our corpus,
we consider this is mainly due to another advantage of \DNFF{}---enhancing the connection between the current and following events
to \textbf{emphasize the causality between them}.

%% T + ZZ
\begin{wrapfigure}[2]{l}{0.13\textwidth}
  \begin{center}
    \vspace{-12pt}
    \includegraphics[width=0.04\textwidth]{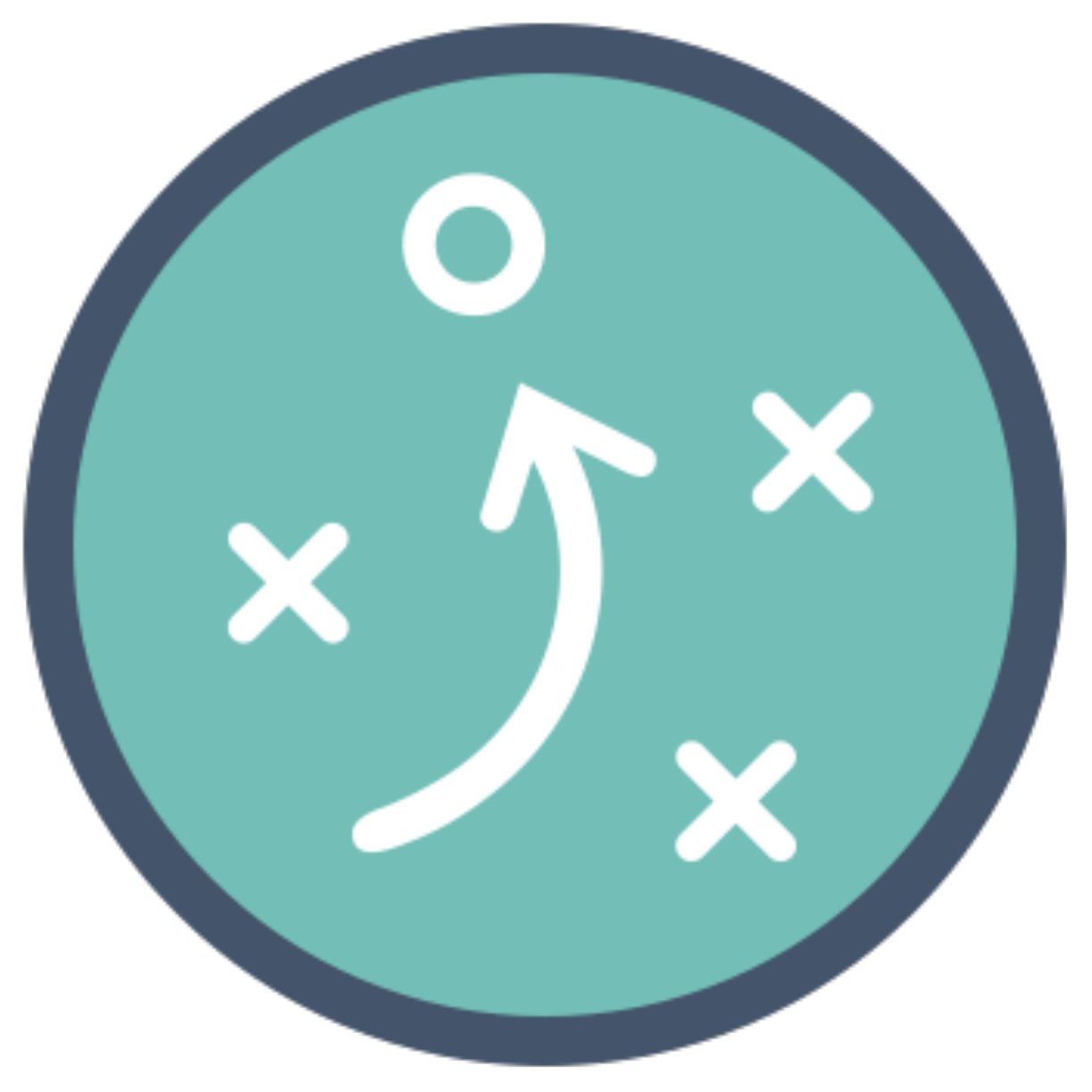}
    \includegraphics[width=0.03\textwidth]{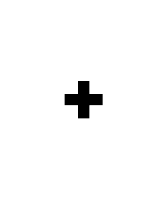}    
    \includegraphics[width=0.05\textwidth]{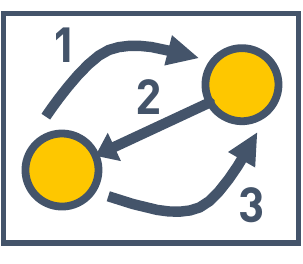}
  \end{center}
\end{wrapfigure} 

\noindent \emph{\underline{\DDC{} + \DNZZ{}.}}
    \DNZZ{} can be used to \textbf{augment one event from different angles}
    % show the data of one event from different angles 
    since it can repeat a specific event multiple times.
    For example, a video can highlight a player in an event 
    and then use \DNZZ{} to replay the same event but with another player highlighted.
    Besides, \DNZZ{} can be seen as \textbf{a composition of two ``single'' narrative orders}.
    For instance, some examples use \DNL{} to show certain data
    % , play the video,
    and then use \DNZZ{} to roll back and show another data using \DNFF{}.
    Given these characteristics, \DNZZ{} has been frequently used to present tactic-level data.

\setlength{\intextsep}{12.0pt plus 2.0pt minus 2.0pt}
\setlength{\columnsep}{10pt}
% \clearpage
% \setlength{\intextsep}{12.0pt plus 2.0pt minus 2.0pt}

\section{\system{}}
Based on the design space,
we design \system{}, 
a proof-of-concept system that 
augments the videos for table tennis.
While the implementation of \system{} is specific to table tennis, 
its design can be generalized to other racket ball sports
once the underlying dataset and ML models are available.

\subsection{Design Goals}

% Based on our collaboration with sports analysts who provide
% data analysis services for national sports teams, we are informed that
% augmenting videos are particularly useful for sports analysts, allowing
% them to effectively communicate the analytical insights to the coaching
% team and players. However, sports analysts usually cannot create these
% videos as the creation process involves complex design decisions and
% video editing. Thus, fast prototyping tools that allow sports analysts to
% create augmented videos rapidly are in great demand.

We conceived the design goals of \system{} 
based on the design space, 
prior research~\cite{WangSZCXMZ20, kim2020, shicalliope} on intelligent data visualization design tools,
and weekly meetings with two domain experts over the past eleven months.
One of the experts is a sports science professor who
provides data analysis and consultancy services for national table tennis teams.
The other is a Ph.D. candidate in sports science with a focus on table tennis analytics.
In the meetings, we demonstrated several prototype systems
and collected feedback to refine our design goals.

\textbf{G1. Extract the Data from the Video based on Data Levels.}
According to our design space, 
augmented sports videos usually present data from different levels (\emph{i.e.}, object-, event-, and tactic-levels)
for entertainment or education purposes.
However, manually preparing the data of different levels might be tedious and time-consuming.
Thus,
to support these augmentations, % and reduce the analyst's workload, 
% The analyst may augment the video with different data for different purposes.
the system should automatically extract and organize the data from a video based on the data levels.
% \cmo{Besides, the data should be organized in a hierarchical manner according to the data level
% for efficiently accessing.}

% \textbf{G2. Augment the Video with Data Instead of Graphical Marks.}
\cmo{\textbf{G2. Interact with the Data Instead of Graphical Marks.}}
The fundamental goal of sports analysts is to visualize the data, such as showing the ball trajectory.
% \emph{e.g.}, show the ball trajectory. 
However, 
most of the video editing tools support the user to augment the raw footage with graphical marks,
such as drawing a line in the scene.
% \emph{e.g.}, draw a line in the scene.
Hence, the system should allow the analysts
to select the data to visualize instead of asking them to select the graphical mark to draw. 
Furthermore, the user should be able to directly interact with the data in the video where it is originated from.

% \noindent
\textbf{G3. Recommend Visualizations for Different Narrative Orders.}
Visualization recommendation systems can ease the creation process as the sports analysts may lack sufficient
knowledge of data visualization.
Prior systems~\cite{voyager, draco} usually recommend  visualizations
based on the effectiveness of visual channels.
However, in our scenarios, 
the narrative order may result in different availability of visual effects.
For example, when using a \DNL{} order, we cannot move a player in the video out of his position.
Consequently, the system should further consider the narrative order
in the recommendation.

\begin{figure*}[hbt]
  \centering
  \includegraphics[width=0.99\textwidth]{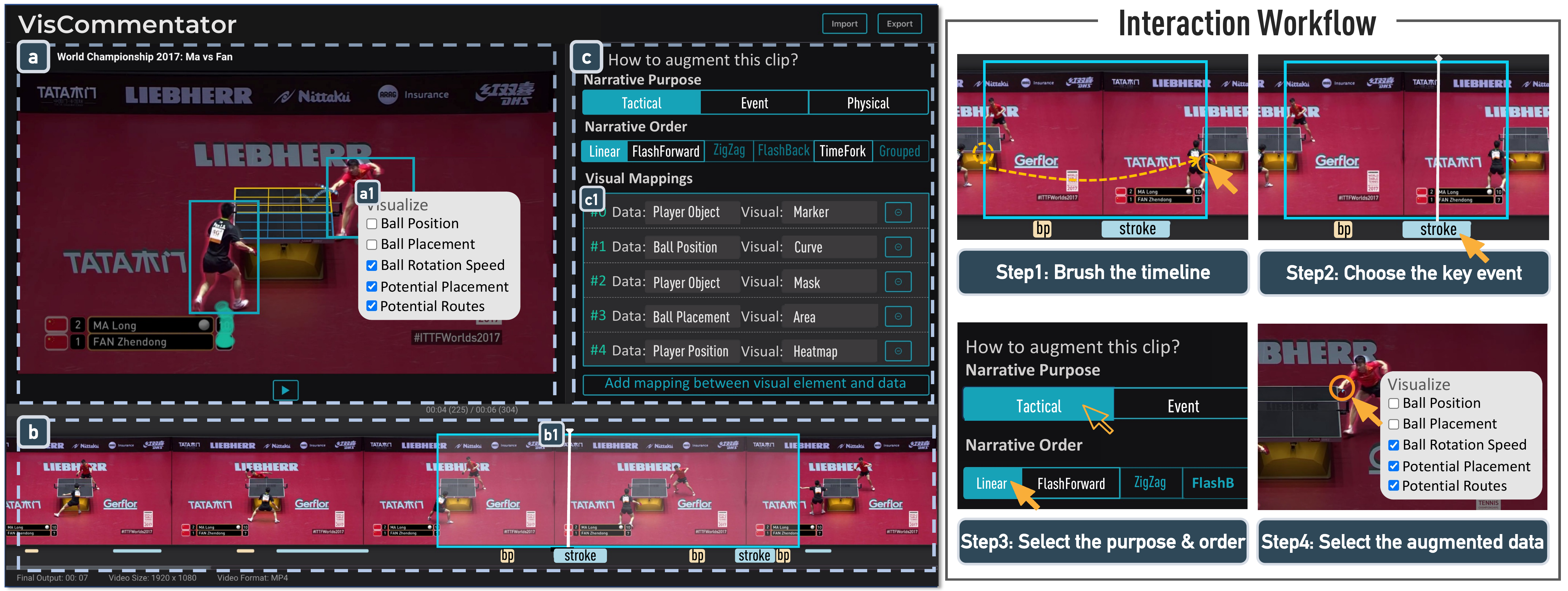}
  \vspace{-2mm}
  \caption{
    The user interface of the system, including a) a video preview, b) a timeline,
    and c) an edit panel. 
    \cmo{A basic authoring workflow includes four steps: 1) brush the timeline, 2) choose the key event,
    3) select the Narrative Purpose and Narrative Order, and 4) select the augmented data by directly interacting with the video objects.
    % The user can directly interact with the objects in the video (a1) to select the data to augment the video. 
    The system will suggest the visuals for the selected data in the edit panel (c1).}
  }
  \vspace{-4mm}
 \label{fig:ui}
\end{figure*}

\subsection{Usage Scenario}
To illustrate how \system{} encompasses the design goals,
we introduce the workflow to create an augmented video taken by Jon,
a hypothetical sports analyst
who needs to analyze a table tennis rally
and present his findings to the coaching team and athletes.
He also needs to create a video with highlight moments to engage the team fans.

\para{Brush the timeline.}
Jon plans to present his analytical findings that
the last two turns are the key events,
in which the ball rotation speed is so fast that the player in red cloth ($Player_{red}$)
can only return it to a narrow area and thus gives his opponent a chance to attack.
Thus, Jon loads the raw video to \system{}.
It's user interface (UI) is familiar to Jon 
as it follows the design of general video editing tools,
which usually have a video preview (\autoref{fig:ui}a),
a main video timeline to be edited (\autoref{fig:ui}b),
and an edit panel (with tactic-level data and Linear order in default) in the right-hand side (\autoref{fig:ui}c).
The events of the rally are automatically detected and visualized under the timeline (G1).
Jon brushes on the timeline to select the last two turns. % to augment the video.

\begin{figure}[hbt]
  \centering
  \includegraphics[width=0.9\columnwidth]{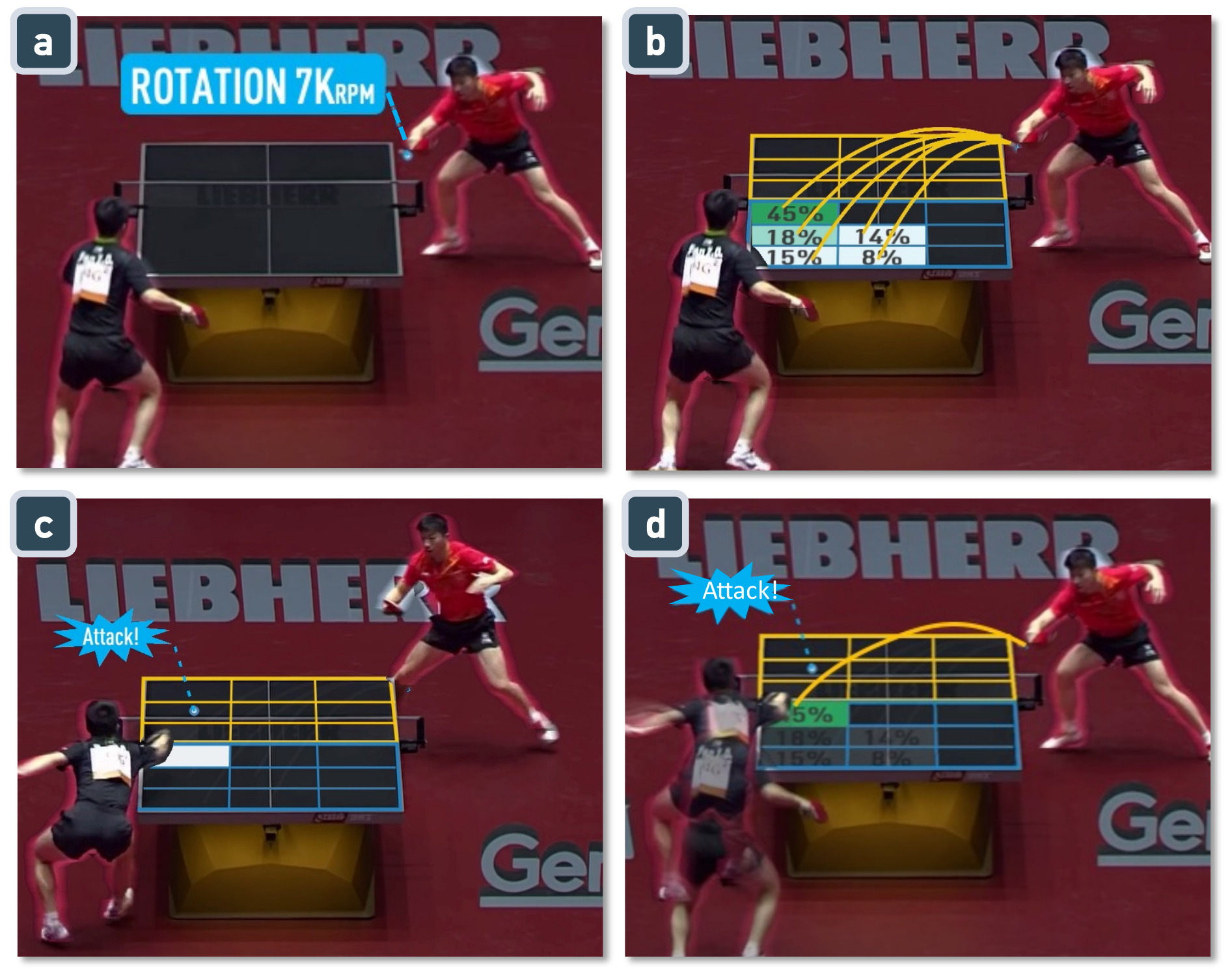}
  \vspace{-2mm}
  \caption{
    a) The rotation speed of the ball is 7000 rounds per minute.
    b) The probability distribution of the ball placement.
    c) The player attacks the ball to win the rally.
    d) The video uses a \DNFF{} to preview the action of the player after showing the potential ball placement.
  }
  \vspace{-6mm}
 \label{fig:scenario}
\end{figure}

\para{Select the data.}
The two players and ball are automatically detected and highlighted 
by using bounding boxes in the main view (\autoref{fig:ui}a, G1).
Jon first moves the pointer to the frame when $Player_{red}$ performs the stroke (\autoref{fig:ui}b1)
and right-clicks the ball in the main view.
A context menu popups,
allowing Jon to directly select which data to be augmented (\autoref{fig:ui}a1, G2).
Based on his analysis,
Jon selects the ``Ball rotation speed'', ``Potential placements'', and ``Potential routes''
in the context menu.
The system automatically recommends visuals for these data 
and lists the mappings in the ``Visual Mapping'' list (G3).
By playing the clip,
Jon finds that the clip pauses and visualizes 
the ball rotation speed by using a label (\autoref{fig:scenario}a),
followed by an animation that shows 
the potential routes and placements possibility of the ball (\autoref{fig:scenario}b).
% Interested in how the augmentation looks like, 
% Jon plays the video and finds out that the video is augmented by visualizations.
% The augmented clip pauses and visualizes 
Satisfied with these augmentations,
Jon moves the pointer to the next turn,
right-clicks the player in black cloth ($Player_{black}$),
and selects the ``Stroke Technique'' data.
Consequently, \system{} augments the video to highlight the active attack of $Player_{black}$ (\autoref{fig:scenario}c).
Although impressed with these augmentations, 
Jon is curious about the ``Narrative Order'' and chooses ``Flash Forward'' 
to check the augmentation effects.
The system presents another augmented clip that
foreshadows the action of the $Player_{black}$ in the video
right after showing the probability distributions of the ball placement
(\autoref{fig:scenario}d).
Jon recognizes that this narrative method 
explains the causality between the ball placement and the $Player_{black}$'s attack in a stronger manner. 

% \vspace{-1mm}
\para{Edit and Fine-tune.}
Besides the coaching team and athletes, 
Jon also needs to present the highlight moments to engage the team fans.
Jon thinks that the tactical data is too complex for the general fans.
Thus, he chooses the \emph{object-level} data in the edit panel to augment the video.
% \emph{Engagement} for the ``Narrative Purpose'',
% which allows him to augment the video with physical data.
Instead of annotating the players' key movements,
Jon now can select some physical data to augment the video.
% Jon chooses to highlight the players' pace with slow motion (\autoref{fig:ui}a).
% the resulting augmentation highlights the ball trace with slow motion (\autoref{fig:ui}a),
% which Jon believes will be favored by the audiences.
He thus directly interacts with the objects in the video to select 
% three more object-level data (\emph{i.e.},
``Player Object'', ``Player Position'', ``Ball Trajectory'', and ``Ball Placement".
% )
% through with the objects in the video.
% in the ``Visual Mapping'' list to augment the clip.
\system{} automatically recommends visual effects to present these data, 
so that Jon only needs to perform minimal editions, \emph{e.g.}, specify the colors.
Satisfied with the results, Jon exports the augmented video.

\begin{figure*}[th]
  \centering
  \includegraphics[width=0.99\textwidth]{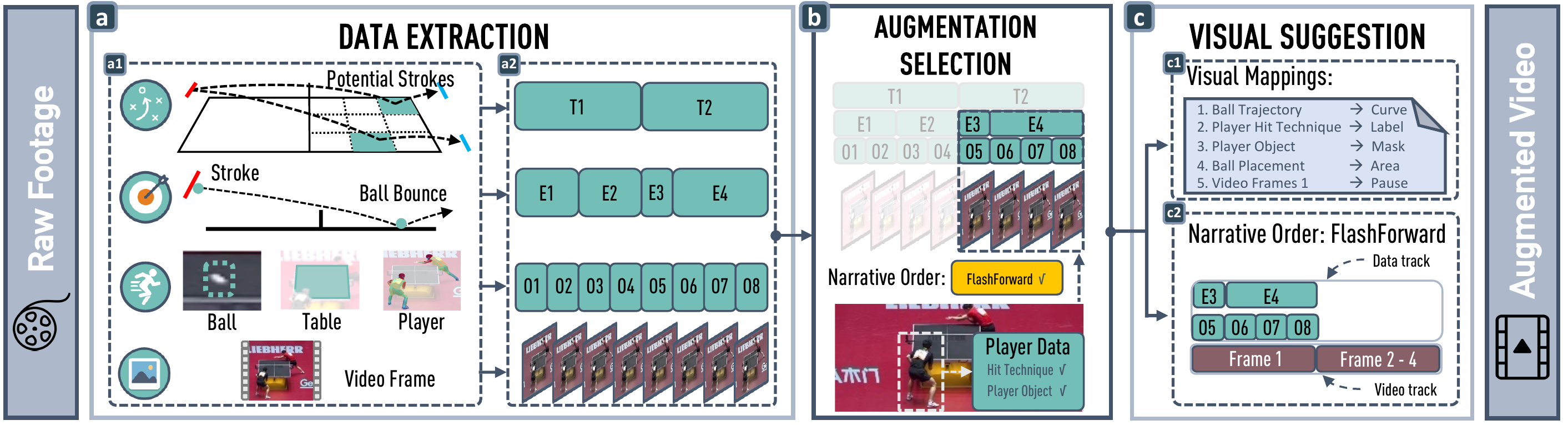}
  \vspace{-2mm}
  \caption{
  The pipeline to augment a raw footage. 
  a) The system (1) adopts a bottom-up approach to extract object-, event-, and tactic-levels data from the video frames,
  and (2) organizes them into a hierarchical manner.
  b) The user can directly interact with the objects in the video to select the data. 
    The narrative order will affect how the data is rendered.
  c) The system automatically (1) suggests visuals for the data and (2) schedules their rendering relatively to the video frames based on the narrative order. The data and video frames are rendered in two separated tracks.
%   c) After selecting a level of data and brushing a specific period, 
%   the system returns a subtree and automatically suggests interesting data insights, \emph{e.g.}, the two events in cyan.
%   d) The system constructs a Directed Acyclic Graph based on the user-selected
%   narrative orders and suggests visual mappings for each data nodes.
%   f) Rendering and generating the augmented video. 
  }
 \label{fig:pipeline}
  \vspace{-4mm}
\end{figure*}

\subsection{Processing the Data through ML Models}
% Based on our design space, the analyst may augment the video for different purposes.
% To support these augmentation,
% \system{} needs to obtain
% the data from different levels, namely, \DDO{}, \DDE{}, and \DDC{}.
% Usually, the data is prepared and imported from other dedicated data analytic tools~\cite{}.
% To ease the workload,
To incorporate G1, we implement a bottom-up approach 
to extract the data from table tennis videos (\autoref{fig:pipeline}a1).
% The analyst can also import the data from other tools~\cite{Deng2021}.
% Considering the limitations of ML models,
% we also allow the user to import the data from other tools~\cite{Deng2021}.

%% 2.1 object level
\para{Object-level Data.}
%% what object level data
We assemble multiple state-of-the-art 
deep learning-based models to extract object-level data from the input video.
Specifically, in each frame, we aim to detect the positions of the ball, the table, and the two players. For the players, we further extract their postures.
To achieve these goals, 
we first extract the feature map of each frame by using a ImageNet~\cite{Deng2009} pre-trained RestNet-50~\cite{he2016} backbone.
The feature map is used for the following detections:
\begin{itemize}[leftmargin=*]
    \item To detect the ball, 
    % multiple models targeted for racket sports are available~\cite{ttnet, huang2019}.
    % In this work, 
    we use TTNet~\cite{ttnet}, a multi-tasks model that can detect and segment the ball, table, and players, as well as classifying ball events such as ball bounce and net hit.
    We only use TTNet to detect the ball and table since TTNet cannot detect player postures.
    Given that the ball is small and moves fast, 
    TTNet employs a two-scales architecture on a stack of consecutive frames to detect the ball.
    The detection is represented by a bounding box.
    Besides, we interpolate the ball position when it is occluded.
    \item The table can be easily detected as it is usually fixed in the video.
    \item For the players, we adopt BodyPix~\cite{bodypiex2} to detect the two players, segment their pixels from the raw image, and detect their postures. BodyPix is an industry-level model for real-time human body pose estimation. 
    The output for each player is represented as the bounding box, the pixels, and the posture key points.
\end{itemize}
Finally, based on these data, we can further calculate other object-level data, 
such as ball velocity, ball trajectory, player moving direction.
Given the ``off-the-shelf'' nature of these models,
we only tested our approach in a 6-seconds long table tennis video, 
whose resolution is $1920 \times 1080$ and frame rate is 50FPS,
and obtained $90\%+$ average precision with intersection over union equals 0.5 for all the objects. 

%% 2.2 event level
\para{Event-level Data.}
% Event-level data is associated with events. 
An event usually covers multiple frames. % and the corresponding object-level data.
Based on the object-level data, 
we extract two types of events, \emph{i.e.}, ball, player: % in a table tennis rally:
\begin{itemize}[leftmargin=*]
    \item Ball events include ball bounce and net hit. We reuse the event spotting branch integrated in TTNet to classify whether there is a ball bounce or net hit event within in a frame. 
    \item The main player event is stroking. 
    To detect the stroke event, we leverage the ball velocity and the distance between the ball and the player's right hand that is detected as a key point of the player's posture. 
    Given that the player's right hand sometimes may be miss detected due to occlusion, 
    we also use the player's neck (another key point of the posture) as a fallback. 
    We \cmo{classify a frame as belonging to} a player stroke event when the distance reaches a bottom 
    and flag it as a ``hit'' frame if the ball velocity \cmo{changes direction}.
\end{itemize}
Finally, based on the events, we further compute other event-level data,
such as the placement region based on the ball position in bounce event, 
the stroke technique based on the player posture in the hitting frame.

%% 2.3 tactical level
\para{Tactic-level Data.}
Tactic-level data usually involves multiple events and deep domain knowledge
% (\emph{e.g.}, player tactic, stroke effect, potential placements),
and thus can hardly be detected from the video by using CV models.
Instead, we adopt a rule-based method to acquire some kinds of tactic-level data.
Specifically, we use a set of rules provided by our domain experts 
to infer the player tactic, 
potential ball routes and placements,
and stroke effect of a player.
% based on the previous and current stroke technique.
\cmo{For example, given the rule that \emph{``if a player received the ball at his end line, he can only return the ball to the end line of the opponent''}, we can calculate the potential ball placements based on the event-level data.}
% uses an offensive technique,
% the opponent can only use a defensive technique or an offensive one to return the ball.}
In addition, we use a pre-trained causal graph~\cite{Causality, DBLP:journals/tvcg/XieDW21} to detect the key stroke that contributes most to the rally result (\emph{i.e.}, why one player win).
Nevertheless, 
there are still many kinds of tactic-level data that cannot be automatically obtained from the video
due to the limitation of the state-of-the-art techniques.
As a trade-off, our system allows the analyst to import external data from other tools~\cite{Deng2021, DBLP:journals/tvcg/WangWCZZW21}.
Further technical details can be found in the supplemental materials.
% ~\footnote{\url{https://viscommentator.github.io}}.

% \para{Data Organization.}
% We organize the extracted data in a tree-based structure (\autoref{fig:pipeline}a2)
% that contains four levels, each corresponding to a data level.
% Each node stores its own data, 
% a pointer to its children,
% and a pointer to its successor.
% By this, we can separate the data at different levels
% while maintaining their chronological orders
% and relationships (\emph{e.g.}, a ball event \emph{net hit} and its physical data \emph{positions}).

% \subsection{Augmenting the Video through Direct Manipulation}

\subsection{\cmo{Interacting with the Data through Direct Manipulation}}
Different from general video editing,
the ultimate goal of sports analysts is to augment the video with data rather than graphical marks (G2).
To this end, a straightforward method is to allow the analyst to select the data in the edit panel (\autoref{fig:ui}c1). 
However, this method leads to a large distance~\cite{hutchins1985direct} between the user's intentions and the meaning of expressions in the interface.
\cmo{For example,} the analyst wants to select the data of the objects in the video, but she has to interact with the edit panel.
% between the data items in the list and objects in the video, 
% thereby increasing the analyst's cognitive load.
To reduce the analyst's cognitive load,
we leverage the extracted data to decorate the video, thereby increasing the directness of interactions. Specifically,
we visualize the events along with the timeline and encode their event type (\emph{i.e.}, ball or player event) with color, so that the user can quickly identify and navigate to a specific turn in the rally.
Besides, the players and ball in the main view (\autoref{fig:ui}a) are right-clickable,
by which the user can directly select the data for augmentation.
% (\emph{e.g.}, ball trajectory, player actions).
All the selections will be mapped to the underlying data (\autoref{fig:pipeline}b) for mapping to visuals.

\para{Fine-tuning in the edit panel.}
Meanwhile, all the selected data will be listed in the edit panel (\autoref{fig:ui}c1)
and encoded with the visuals suggested by the system.
The user can further adjust and fine-tune the visual effects, such as 
modifying the color, line width.

% After extracting the data from the input video,
% based on our design space, 
% the user need to specify the data level, narrative order, and the visual-data mapping to augment the clip.
% The data level and narrative order can be simply selected in the edit panel.
% The selection result will affect what and how data is used to augment the video.
% For the visual data-mapping, 
% our system provides a set of intuitive interactions (G2):
% to augment the video (G2).

% \para{Direct manipulation in the video.}
% Compared with mainstream video editor tools,
% our system decorates the video with the extracted data to increase the directness of interactions~\cite{hutchins1985direct}:

\subsection{Visualizing the Data through Recommendation}
% Based on the data level, narrative order, and data selected by the user,
\system{} integrates a visualization recommendation
engine to map the data to visuals based on the user-selected narrative order (G3).
% different visual effects
% and plan the appearance orders of the augmentations (G3).

\para{Visuals---Maximum Conditional Probability.}
% Given the data selected by the user, the system should suggest proper visuals.
% There are many techniques 
% for mapping data to visuals,
Previous research proposed several methods to suggest visuals based on the data,
such as rule-based~\cite{voyager}, constraint-based~\cite{draco}, decision tree~\cite{WangSZCXMZ20}, and neural network~\cite{viznet}.
Considering our specific domain, 
we intend to use the collected videos as prior knowledge to recommend the visuals.
Specifically, 
we model the visual mappings using conditional probability distribution: $p = f((d, v) \mid O)$,
where $d$, $v$, and $O$ is the data, visual, and narrative order, respectively.
Intuitively, this model represents the probability distribution of data-visual mappings under different narrative orders. 
This probability distribution can be estimated based on the occurrence frequency of the combinations of data, visual, and narrative orders in our corpus.
Consequently, given $d$ and $O$, 
our system will search the $v$ that maximizes $p$ and suggest it to the user.
We also use a rule-based method~\cite{voyager} as the default 
to select effective visual channels for the data
if there are no mapping records of the data under a narrative order.
Our method is simple but effective 
since it is built on videos 
% collected 
from reputable sources,
and can be extended and optimized 
in the future.

Finally, since the position of the objects in each frame are detected,
we can easily render the visual effects of the objects in the screen space.

\para{Narrative Order---Double-track Rendering.}
To correctly rendering the visuals 
based on the user-selected narrative order,
we employ a double-track rendering method (\autoref{fig:pipeline}c2).
Specifically, we render the frames and data visualizations in two parallel tracks (video track and data track)
to control their appearance order.
% by using two independent timelines.
For example,
in \autoref{fig:pipeline}b, the user selects the data of four frames 
and renders them by using a FlashForward, 
\cmo{which presents the data that will happen later than the current frame.}
Our system will pause the video track after \cmo{playing} the first frame
and keep rendering the data for the next three frames.
The video track will be resumed after rendering all the selected data.
% After visualizing all the selected data,
% the video track will be resumed.
% The next task is to decide the appearance order of the data in relation to the raw video.
% In \system{}, we model the appearance order of different data
% by using a Directed Acyclic Graph (DAG).
% The DAG is constructed based on the \DN{} selected by the user.
% For example, 
% in \autoref{fig:pipeline}c,
% if the user selects a \DNFF{},
% our system will automatically add a virtual link between node $E3$ and $E4$ (\cmo{the dashed link in \autoref{fig:pipeline}d}).
% In the actual execution of the animation,
% Then when executing the animation,
% our system will perform a topological sort to ensure
% all the prepended nodes of $E3$ are presented before animating to $E4$.
% Besides, $E4$ will be flagged to avoid repeated visualizations.
The duration of the visualizations in a non-linear order is set with a default value.
The current system only implements the common patterns mentioned in Sec.~\ref{ssec:common}.

% Obviously, not in all cases, a virtual link should be added between two nodes.
% The addition of links between two nodes depends on their data attributes to show.
% % For example, \cmo{what example?}
% In our implementation, 
% we use the annotations of our video corpus as prior knowledge
% to decide whether a link should be added or not.
% Specifically, 
% before adding virtual links between two nodes 
% (\emph{e.g.}, use \DNFF{} to show a player's position in $E4$ 
% after showing the placement of a ball in $E3$),
% we will search the corpus to ensure such pattern exists; 
% otherwise, the link will not be added.
% In current version, we only implemented the common patterns mentioned in Sec.~\ref{ssec:common}.

\section{Implementation}

\system{} is implemented in a browser/server architecture.
The browser part, which is built upon HTML + CSS + JavaScript, 
is responsible for the UI and rendering of the video.
We mainly use HTML Canvas to achieve the rendering of augmented videos.
To improve the web-page efficiency,
we use the OffscreenCanvas~\cite{OffscreenCanvas} functionality that leverages
the \emph{worker}, a multi-thread like technique in a modern browser,
to accelerate heavy rendering.
The server part is implemented based on Node.js + TypeScript.
To extract the data from a video,
we employ PyTorch~\cite{pytorch} and 
TensorFlow.js~\cite{tfjs} that support
running pre-trained deep learning models by using Node.js.
% The project will be made open-source once the paper is accepted.
% wherein the browser part is responsible for the UI
% and the server part handles the data processing.

\begin{figure*}[h]
  \centering
  \includegraphics[width=0.99\textwidth]{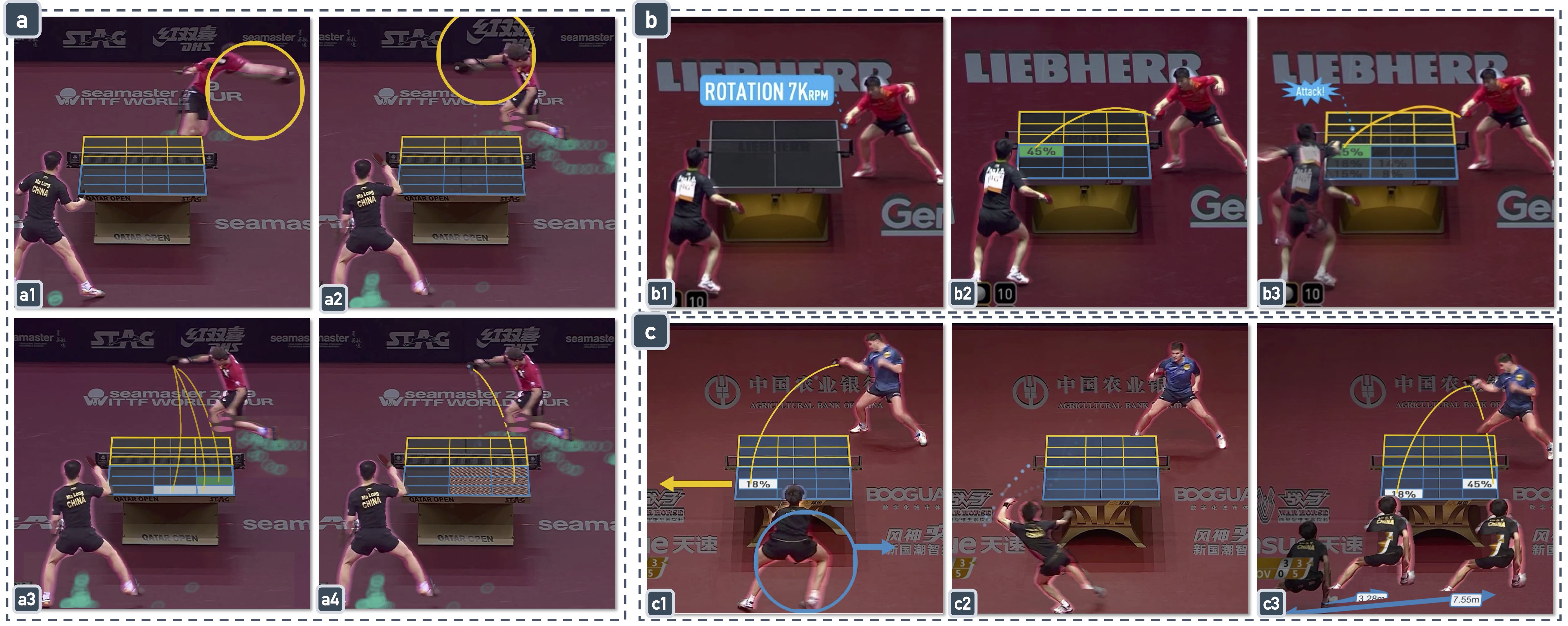}
  \vspace{-2mm}
  \caption{
    a) The video highlights that (1) the player in red first uses his left hand.
    (2) He switches to his right hand to stroke the ball to (3) obtain more chances to hit the ball to the corner, and (4) thus successfully hits the ball to the empty area.
    b) The video visualizes (1) the ball rotation speed and foreshadows (2) the ball route and placement, as well as (3) the player's next movement.
    % The video first shows the Tactical level data that reveals the player in red strokes the ball to an empty area that his opponent cannot cover.
    % f) Then the video rollbacks to the previous turn and replays again.
    % g) In this time, the video 
    % h) However, In the next movement, 
    c) The augmentation shows that (1) when the player in black plans to move to the right, his opponent strokes the ball to the left.
    (2) The player in black tries his best to return the ball.
    (3) After returning the ball, his opponent can stroke the ball to the two table corners. The opponent strokes to the farther corner and win the rally. 
    % the ball to the right, which is too far away for him to return the ball. Thus, he loses this rally.
  }
 \label{fig:case}
 \vspace{-4mm}
\end{figure*}

\section{User Study}

We conducted a \cmo{qualitative} user study to assess the usability of \system{}. 
The study aimed to evaluate whether sports analysts, the target users,
can create augmented table tennis videos with our system
and to observe their creation process to reflect on future improvements.
% We did not run a comparative study with, \emph{e.g.}, Adobe Premiere, as a baseline
% since the features of our system do not exist in Premiere.

\para{Participants}:
We recruited 7 table tennis analysts (P1-P7; 4 male; age: 20-30, one didn't disclose)
from a university sports science department.
All experts majored in Sports Training 
with proficient experience in analyzing table tennis matches.
All the experts had the experience of using lightweight video editors, \emph{e.g.}, VUE~\cite{vue}, but no experience on advanced video editing tools such as Adobe Premiere and Camtasia.
The scale of expert participants is consistent with similar sports visualization research~\cite{wu2017ittvis, stein2018}.
Each participant received a gift card worth \$14 at the beginning of the session, independent of their performance. 

\para{Tasks}:
The participants were asked to finish a training task and two reproduction tasks by using our system.
We prepared three augmented videos
based on three ITTF top 10 rallies in 2019~\cite{ittf}:
\textbf{T1} (\autoref{fig:case}b) is augmented by event-level data and presented by FlashForaward \cmo{with 5 visual-data mappings};
\textbf{T2} (\autoref{fig:case}c) is augmented by tactic-level data and presented by TimeFork \cmo{with 10 visual-data mappings};
\textbf{T0} (\autoref{fig:case}a) includes two augmented clips and covers all the features in \textbf{T1} and \textbf{T2}.
All three augmented videos were created by our domain experts in the collaborations.
For the tasks,
we provided the augmented videos,
background information,
the system manual,
a digital version of the design space,
and the data extracted from the video.

% table tennis videos (\emph{i.e.}, \cmo{xxx})
% which were selected as top 10 rallies in 2019 by ITTF~\cite{mavsboll, linvsOvt}.
% The resulted augment videos cover two data levels (\emph{i.e.} object- and tactic-levels) 
% and three most frequently used narrative orders (\emph{i.e.}, Linear, FlashForward, and TimeFork).

% For the training task, 
% we further prepared another augmented video (\autoref{fig:case}\cmo{a}), \textbf{T0},
% which is also created based on a ITTF top 10 rallies and covers all the features in \textbf{T1} and \textbf{T2}.

% \begin{figure}[h]
%   \centering
%   \includegraphics[width=0.48\textwidth]{figs/6_studyprocedure.png}
%   \caption{The overview of the study procedure.}
%  \label{fig:study_procedure}
% \end{figure}

% \vspace{-1mm}
\para{Procedure}: 
% The procedure is summarized in \autoref{fig:study_procedure}.
The study began with the introduction (15min) of 
the study purpose,
the concept of augmented sports videos,
and the design space with 15 curated example videos.
% We also played an official tutorial video of Vizrt Libero~\cite{vizrttutorial},
% a state-of-the-art commercial tool,
% to introduce the typical creation process of augmented sports videos.
We moved to the training phase (20min) when the experts had no more questions.
% of the introduction.
The participants were walked through with a step-by-step instruction to reproduce \textbf{T0}.
% Questions and discussions were encouraged throughout the whole process.

After the training, participants were 
provided with the materials of the two reproduction tasks, \textbf{T1} and \textbf{T2} (15min for each).
We encouraged them to ask questions about the materials.
A task was started when a participant was confident to begin, 
and ended when the participant confirmed finishing. 
To assess the efficiency and effectiveness of the system,
we recorded the completion \emph{time} and \emph{successfulness} for each task.
% A task is failed if the participant cannot reproduce the video in 15 minutes.
A successful reproduction should correctly reproduce 
the data level, narrative order, and all the visual-data mappings of the given video in 15 minutes (\cmo{based on our pilot study}).
All measures were explicitly explained before the tasks.
The participants were encouraged to explore the system (20min) after the tasks.

The session ended with a semi-structured interview and a post-study questionnaire (15min).
Each session was run in the lab, using a 24-inch monitor, 
with a think-aloud protocol, and lasted around 90 minutes.
% 7-point Likert scale
% The correct points of \textbf{T1} and \textbf{T2} are 7 and 11, respectively,
% including 

\subsection{Results}
\cmo{We consider the study assessed the system qualitatively rather than quantitatively due to the small sample size.}
All participants could successfully reproduce the two augmented videos in a few minutes
\cmo{
  (\textbf{T1}: $~$5mins; 
  \textbf{T2}: $~$9mins).}
% (\textbf{T1}: $\mu = 322s, 95\%~CI = [203, 440]$; 
% \textbf{T2}: $\mu = 526s, 95\%~CI = [359, 693]$).
\cmo{
Some participants (P1, P3, P4) could not memorize the details of the targeted videos 
and thus frequently re-played the videos to check the augmentations during the authoring. 
This is the reason why these participants took a longer time to finish the tasks.}
% We did observe that some participants (P1, P3, P4) could not correctly reproduce
% the videos in the first place. 
% But they could iteratively refine the results for the correct one by checking the targeted output.
% This is also the reason why these participants took a longer time to finish the tasks.
Besides, we also noticed that P2 and P5 explored other augmentations during the tasks
even they had already successfully created the results, leading to extra time to finish the tasks.
Overall, these results \cmo{qualitatively} demonstrated the effectiveness and efficiency of our system.
% We also collected and summarized 
The participants' feedback is summarized as follows:

% \begin{figure}[h]
%   \vspace{-1mm}
%   \centering
%   \includegraphics[width=0.49\textwidth]{figs/62_usability_chart.png}
%   \caption{
%     Participants' qualitative feedback on the system, showing means and 95\% CIs. 
%     Distributions are shown on the left.
%   }
%  \label{fig:author_rating}
%  \vspace{-5mm}
% \end{figure}

\para{Usability}: Overall, our system was rated as \emph{easy to learn} 
\cmo{($\mu = 6.00$)}
% ($\mu = 6.00, 95\%~CI = [5.14, 6.86]$)
and \emph{easy to use} 
% ($\mu = 6.86, 95\%~CI = [6.58, 7.14] $) 
\cmo{($\mu = 6.86$)}
by the participants.
The participants (P1-4, P6) commented that \emph{``...this system allows me to do more with less effort.''} 
P3 provided in detail that \emph{``I have used other video editors...I can't imagine how I can use them to create augmented videos...''}
When we asked the participants whether they need more customization of the augmentations 
such as the order of visuals,
most of them agreed that on one hand, the tool should 
strike a balance between customizability and simplicity
as their goal is \emph{``rapid prototyping''};
on the other hand, the tool should allow editing of the details when the user needs to.
% allow controlling the details
% which part is the most difficult to learn,
% four of the participants thought was to \emph{``''}
% two (P3 and P7) 
% As commented by P3, \emph{``the system doesn't have too much complex UI''} and
% \emph{``I can quickly generate augmentations by clicking several buttons.''}
% Although we had never mentioned the concept of templates,
% P4 pointed out that \emph{``your system applies templates to the video...''}.
% When we asked which part is the most difficult to learn, 
% four of the experts thought was to \emph{``select proper visual effects''},
% two (E4 and E5) said the concept of narrative order,
% and one (E1) believed nothing was too difficult.

\para{Usefulness}:
The key designs---data extractor, direct manipulation UI, and visuals recommendation--of our
% The data-driven features---data insights and visuals recommendations---of our
system were particularly lauded by the experts 
and confirmed to be 
useful \cmo{($\mu = 6.29$, $\mu = 6.86$, and $\mu = 6.57$, respectively).}
% ($\mu = 6.29, 95\%~CI = [5.73, 6.85]$, $\mu = 6.86, 95\%~CI = [6.58, 7.14]$, and $\mu = 6.57, 95\%~CI = [6.00, 7.15]$, respectively).
The participants thought that extracting and organizing data based on data levels were useful.
P2 commented that \emph{``organizing data into the three levels is not surprised to me, but connecting them to different narrative purposes inspires me a lot.''}
All the participants spoke highly of how we support the augmentation through direct manipulation.
\emph{``Clearly revealing the object to be augmented''} is the main reason why the participants favored it.
P7 further pointed out that the UI design made the task \emph{``cognitively easy.''}
The participants found the visualization recommendation was particularly useful
as they \emph{``were not familiar with mapping data to visuals.''}
% \emph{``Although I can do that by trial and error, it is time consuming.''}
Besides, P5 suggested allowing \emph{``comparison between different recommendations.''}

\para{Satisfaction}:
The rating also reflected a high user satisfaction for the data extractor
\cmo{($\mu = 6.29$)},
% ($\mu = 6.29, 95\%~CI = [5.36, 7.20]$),
direct manipulation design 
% ($\mu = 6.43, 95\%~CI = [5.85, 7.00]$),
\cmo{($\mu = 6.43$)}
and visualization recommendation 
\cmo{($\mu = 6.43$)}.
% ($\mu = 6.43, 95\%~CI = [5.85, 7.00]$).
The participants said that the direct manipulation design was \emph{``intuitive''} (P1-P7)
and the visualization recommendation \emph{``helps guys like us (sports analysts)...''} (P1, P6)
Comments also implied further improvements of the data extractor:
\emph{``The available tactic data is a bit limited..''} (P3, P6)
We explained that this was mainly due to the limitations of ML models
and our system allows the user to import external tactic data.

\subsection{Observations and Feedback}
% We reflect on the observations of the study and interview feedback.
% which implies future opportunities.

% \para{The benefits of presenting analytical insights through augmented videos.}
\para{Increasing the player's memorability in a tense game.}
We were interested in how the participants usually presented their analytical findings 
and the advantages of using augmented videos.
% to present analytical findings
% compared to the methods that the participants usually use.
% how the participants usually presented their analytical insights of a match
% and what are the advantages of using augmented videos.
All the participants responded that voiceover on the video is the most common way to present their findings.
They also prepared slides to add annotations to the video.
Some participants (P3, P5, P6) also created simple charts, such as bar and pie charts, 
to visualize the data.
Compared to these methods, a particular benefit of augmented videos %, besides \emph{``intuitive.''} (P1-P7),
is \emph{``increasing the memorability.''} (P1, P3-P7)
This is extremely important for the players.
As pointed out by P5,
the game is so tense that
the player usually can only
\emph{``recall the visual scenes rather than the numbers.''}

% which is particularly useful for the players,
% since  (P5) 
% is \emph{``intuitive.''} (P1-P7)
% Specifically, the augmented videos can 
% reduce the cognitive load of the audience to understand and connect the complex data to the physical context.
% P2 explained that \emph{``the audiences do not need to imagine and can directly see what I am talking in the video.''}
% Another advantage is \emph{``(the augmented video) can increase the memorability.''} (P1, P3-P7),

\para{\cmo{Direct manipulation with the video objects.}}
\cmo{While our system allows the user to directly interact with the video elements to select the 
data, we observed that the participants wanted to achieve more than that.
For example, P3 and P7 wanted to navigate the video by directly moving the player;
P2 tried to pan the video to adjust the camera.
These behaviors revealed that the users not only directly interact with the video elements 
to manipulate the data but also the \emph{time} and \emph{space}.
Prior research~\cite{directVideo1,directVideo2} proposed techniques to allow users to navigate the 
video by directly moving the objects. 
DimpVis~\cite{DimpVis} applies the similar idea for time-varying visualizations.
Yet, how to unify the direct manipulation of time, space, and data in a data-augmented video remains unclear. Further exploration of the design space is thus suggested.
}

\para{Bridging the gap between analyzing and communicating data.}
While visual analytics systems have been increasingly used for data analysis,
it remains challenging for analysts 
% to create visual designs 
to visually communicate their analytical findings.
For example, although all the participants expressed strong interest in authoring augmented videos, % to present analytical findings,
they had never created augmented videos before 
since \emph{``it is too complicated to create them...''} (P2)
P7 detailed that \emph{``the design tools mainly focus on the visual parts''} 
so that the analyst has to \emph{``manually map the data insights to the visual elements.''}
Thus, all the participants appreciated that our system allows data-driven 
% other than graphics-driven 
design
as \emph{``the ultimate goal of drawing a curve is to present the trajectory data.''} (P7)
Nonetheless, the participants also suggested that our system should be integrated more closely
with visual analytics systems. For example, 
% instead of asking the analyst to configure the visual mappings manually,
the system should allow the analyst to import the analytical findings
and automatically generate augmented videos.
We consider this suggestion as 
an important future direction of bridging the gap between visual analytics and storytelling.

\para{Translating high-level insights into specific data.}
During the study, we observed that a main obstacle for the participants
is to translate a high-level insight into specific data visualization.
For example, the participants could easily perceive that \autoref{fig:case}c1 showcases an \emph{attacking both sides} tactic.
However, to reproduce this clip,
the participants needed to break down this tactic into specific data 
such as \emph{ball trajectory}, \emph{moving direction} of the ball and player.
This ``translation'' process usually led to a trial-and-error situation.
In the interview, 
the participants commented that this process is similar to a reverse analysis 
in which the analyst needs to \emph{``generate data from insights.''} (P7)
Moreover, P1 provided that in some cases, there is no data equivalent of insights, \emph{e.g.}, \emph{``the player is in low spirits.''}
This finding implies future study to translate the analyst's high-level findings into specific data.
% which is also an important component to bridge the gap between visual analytics and storytelling.
A potential solution is to leverage natural language processing techniques, 
which have already been employed in some creativity support tools~\cite{DBLP:conf/chi/KimDAH19, DBLP:conf/chi/LaputDWCALA13}.

\para{Suggestions.}
% The participants also identified some limitations of \system{} and suggested improvements.
Most limitations identified by the participants were related to system engineering maturity.
% as our system is not meant to be a fully functional video editor.
% \emph{e.g.},
For example,
transitions between clips were not supported;
some parameters (\emph{e.g.}, playback rate) were fixed to default values.
Our system also involves certain inherent limitations due to the underlying ML models.
First, the input video needs to cover the whole court with a fixed camera angle.
Although this is the common case for racket sports, 
videos of team sports such as football and basketball do not satisfy this requirement,
which necessitates additional steps to concatenate different views into a panoramic view~\cite{stein2017bring}.
Second, the data extracted from the video is still limited.
For example, 
the server type, hitting area, and other historical data are unavailable.
To mitigate this issue, 
we allow importing data from dedicated data annotation systems (\emph{e.g.}, EventAnchor~\cite{Deng2021}).
Third, the system only supports augmentation in screen space instead of world space.
In other words, our system does not reconstruct the 3D scene of the video 
and thus cannot properly handle z-order and occlusions.

\section{Future work and limitations}

\vspace{-1mm}
\textbf{Significance---communicating data insights through videos.}
After decades of research efforts,
many systems and methodologies have been proposed
to lower the barrier to data visualizations,
allowing data analysts and even general users to communicate data insights through visualizations.
% Theses research systems and methods have been shifted and integrated into commercial
% softwares, such as Power BI~\cite{powerBi} and Tabular~\cite{Tableau}, to allow 
However, 
communicating data through video-based visualizations is still a challenging task~\cite{Thompson2020}.
% and has not been fully explored.
On the other hand, it has become increasingly popular in recent years to
disseminate information through videos (\emph{e.g.}, YouTube, TikTok) as they can convey information in an engaging and intuitive manner.
We explore this direction by investigating augmenting sports videos with data.
Feedback from our user study with sports analysts demonstrates the potential of communicating data insights through videos.
We expect our work can increase interest and inspire future research in this promising direction.

\para{Generalizability---extending from table tennis to other ball sports.}
Although our prototype system focuses on augmenting table tennis videos,
its main design can be extended to other racket sports once the underlying ML models and datasets are available.
% Our prototype system only supports the augmentation of table tennis videos.
% However, it can be easily extended to other turn-taking ball sports  once data is available.
As for team sports (\emph{e.g.}, basketball, soccer),
future studies are required as these sports involve more complex spatio-temporal data~\cite{DBLP:journals/tits/WuWDBXWZDC21, DBLP:journals/tvcg/WengZDMBZXW21} and need more powerful ML models~\cite{DBLP:journals/tvcg/XieWLDCZCW21}.
The design space, which covers both racket and team sports, 
can provide guidance for designing more extensive authoring tools
to augment sports videos of other ball sports.

% As for team ball-sports (\emph{e.g.}, basketball, soccer),
% the 
% they involve more parallel events 
% and thus require a redesign of the data pyramid, which stores the data in a linear manner.
% Nonetheless, the design space can still provide guidance for designing authoring tools
% to create augmented sports videos for other ball sports.

% \vspace{-1mm}
\para{Potentiality---augmenting not only videos but also reality.}
Our research can be a step-stone towards SportsXR~\cite{lin2020sportsxr},
which focuses on using augmented reality (AR) techniques to augment real-world sports activities
with data visualizations, thereby supporting in-situ decision making and engaging sports enthusiasts.
As delineated by Stein et al.~\cite{stein2017bring},
the study of augmenting videos with visualizations is intter-connected
to AR visualization~\cite{DBLP:conf/chi/ChenTWBQ20}. %, especially SportsXR~\cite{lin2020sportsxr},
However, although AR opens up new opportunities 
for both sports analytics and sports watching experience,
it also introduces many challenges that yet to be tackled,
such as scene understanding,
streaming decision making,
and data visualization in 3D real-world canvas~\cite{ShuttleSpace,marvist}.
We believe our work adds to this direction by 
exploring the ways to present data in real-world canvas.

\para{Study limitations.}
Similar to other sports visualization research~\cite{stein2018},
% Although the scale of expert participants in the study is consistent with similar
the sample size of our user study is small since the access to sports experts is naturally limited. 
\cmo{Thus, the study results are considered to be qualitative rather than quantitative.}
The design space is derived from a corpus of a limited number of videos.
% A larger-scale study that involves more videos and collective knowledge (\emph{e.g.}, crowdsourcing) is thus suggested.
Our system only implements the most common patterns in the design space. 
Other less frequently used combinations of data level and narrative order are left for future implementation.
Finally, although the experts and participants were satisfied with the created augmented videos,
we didn't evaluate the videos from the audience's perspective.
Future study is suggested once the system is engineering mature. 

\section{Conclusion}
This work is motivated by the close collaboration with sports analysts 
who have a strong demand to augment sports videos with data.
To ease the creation of augmented sports videos,
% In this work, we aim to ease the creation of augmented sports videos for sports experts.
% To understand the design practices of 
we first systematically review \numVideo{} augmented sports videos collected from reputable sources 
and derive a design space that characterizes augmented sports videos at element-level (\emph{what are the constituents})
and clip-level (\emph{how the constituents are organized}).
Informed by the design space,
we present \system{}, a prototype system that
allows sports analysts to augment
table tennis videos efficiently.
\system{} extracts data from a table tennis video based on data levels,
allows the user to select the augmentation by interacting with the objects in the video,
and suggests visuals for entertainment or education purposes. 
% embeds two recommendation engines to
% suggest data insights and visual effects 
% based on the user-selected narrative purpose (\emph{e.g.}, for education or engagement) 
% and narrative order.
A user study with seven sports analysts confirmed the effectiveness, efficiency, and high satisfaction of the system.
% The study's created videos were found to be informative and engaging by a group of 23 sports fans.
We have also discussed the observations and feedback from the study,
which suggest future research.

\acknowledgments{
We thank Karl Toby Rosenberg for the video narration.
This work was partially supported by NSFC (62072400) and Zhejiang Provincial Natural Science Foundation (LR18F020001) and
the Collaborative Innovation Center of Artificial Intelligence by MOE and Zhejiang Provincial Government (ZJU).}

\bibliographystyle{abbrv-doi}

\bibliography{template}

\begin{thebibliography}{10}

\bibitem{pytorch}
Pytorch.
\newblock \url{https://pytorch.org}.

\bibitem{amini2015}
F.~Amini, N.~H. Riche, B.~Lee, C.~Hurter, and P.~Irani.
\newblock {Understanding Data Videos: Looking at Narrative Visualization
  through the Cinematography Lens}.
\newblock In {\em Proc. of CHI}, pp. 1459--1468. {ACM}, 2015.

\bibitem{Amini2017}
F.~Amini, N.~H. Riche, B.~Lee, A.~Monroy-Hernandez, and P.~Irani.
\newblock {Authoring Data-Driven Videos with DataClips}.
\newblock {\em {IEEE} TVCG}, 23(1):501--510, 2017.

\bibitem{piero}
BBC.
\newblock Piero.
\newblock \url{https://www.bbc.co.uk/rd/projects/piero}, 2020.

\bibitem{brehmer2017}
M.~Brehmer, B.~Lee, B.~Bach, N.~H. Riche, and T.~Munzner.
\newblock {Timelines Revisited: {A} Design Space and Considerations for
  Expressive Storytelling}.
\newblock {\em {IEEE} TVCG}, 23(9):2151--2164, 2017.

\bibitem{cao2020}
R.~Cao, S.~Dey, A.~Cunningham, J.~A. Walsh, R.~T. Smith, J.~E. Zucco, and B.~H.
  Thomas.
\newblock {Examining the Use of Narrative Constructs in Data Videos}.
\newblock {\em Vis. Informatics}, 4(1):8--22, 2020.

\bibitem{courtvision}
Clippers.
\newblock Court vision.
\newblock \url{https://www.clipperscourtvision.com/}, 2020.

\bibitem{cohn2013}
N.~Cohn.
\newblock {Visual Narrative Structure}.
\newblock {\em Cogn. Sci.}, 37(3):413--452, 2013.

\bibitem{Deng2021}
D.~Deng, J.~Wu, J.~Wang, Y.~Wu, X.~Xie, Z.~Zhou, H.~Zhang, X.~Zhang, and Y.~Wu.
\newblock {EventAnchor: Reducing Human Interactions in Event Annotation of
  Racket Sports Videos}.
\newblock In {\em Proc. of {CHI}}. {ACM}, 2021.

\bibitem{Deng2009}
J.~Deng, W.~Dong, R.~Socher, L.~Li, K.~Li, and F.~Li.
\newblock {ImageNet: A Large-scale Hierarchical Image Database}.
\newblock In {\em {Proc.} CVPR}, pp. 248--255. IEEE, 2009.

\bibitem{directVideo1}
P.~Dragicevic, G.~A. Ramos, J.~Bibliowicz, D.~Nowrouzezahrai, R.~Balakrishnan,
  and K.~Singh.
\newblock Video browsing by direct manipulation.
\newblock In M.~Czerwinski, A.~M. Lund, and D.~S. Tan, eds., {\em Proc. of
  CHI}, pp. 237--246. {ACM}, 2008.

\bibitem{espn}
ESPN.
\newblock Detail.
\newblock
  \url{https://www.espn.com/watch/catalog/f48c68af-f980-4fcb-8b59-2a0db01f50cf/_/country/us
  }, 2020.

\bibitem{fischer2019video}
M.~T. Fischer, D.~A. Keim, and M.~Stein.
\newblock {Video-based Analysis of Soccer Matches}.
\newblock In {\em Proc. of International Workshop on Multimedia Content
  Analysis in Sports}, pp. 1--9, 2019.

\bibitem{he2016}
K.~He, X.~Zhang, S.~Ren, and J.~Sun.
\newblock {Deep Residual Learning for Image Recognition}.
\newblock In {\em Proc. of {CVPR}}, pp. 770--778. {IEEE}, 2016.

\bibitem{viznet}
K.~Z. Hu, S.~N.~S. Gaikwad, M.~Hulsebos, M.~A. Bakker, E.~Zgraggen, C.~A.
  Hidalgo, T.~Kraska, G.~Li, A.~Satyanarayan, and {\c{C}}.~Demiralp.
\newblock {VizNet: Towards {A} Large-Scale Visualization Learning and
  Benchmarking Repository}.
\newblock In {\em Proc. of CHI}, p. 662. {ACM}, 2019.

\bibitem{hutchins1985direct}
E.~L. Hutchins, J.~D. Hollan, and D.~A. Norman.
\newblock {Direct manipulation interfaces}.
\newblock {\em Human-computer interaction}, 1(4):311--338, 1985.

\bibitem{ittf}
ITTF.
\newblock Top10 rally in 2019.
\newblock \url{https://youtu.be/0ff3dAt41pU}, 2020.

\bibitem{kim2020}
N.~W. Kim, N.~H. Riche, B.~Bach, G.~Xu, M.~Brehmer, K.~Hinckley, M.~Pahud,
  H.~Xia, M.~J. McGuffin, and H.~Pfister.
\newblock {DataToon: Drawing Dynamic Network Comics With Pen + Touch
  Interaction}.
\newblock In {\em Proc. of CHI}, p. 105. {ACM}, 2019.

\bibitem{DBLP:conf/chi/KimDAH19}
Y.~Kim, M.~Dontcheva, E.~Adar, and J.~Hullman.
\newblock {Vocal Shortcuts for Creative Experts}.
\newblock In {\em Proc. of CHI}, p. 332. {ACM}, 2019.

\bibitem{DimpVis}
B.~Kondo and C.~Collins.
\newblock Dimpvis: Exploring time-varying information visualizations by direct
  manipulation.
\newblock {\em {IEEE} TVCG}, 20(12):2003--2012, 2014.

\bibitem{DBLP:conf/chi/LaputDWCALA13}
G.~Laput, M.~Dontcheva, G.~Wilensky, W.~Chang, A.~Agarwala, J.~Linder, and
  E.~Adar.
\newblock {PixelTone: A Multimodal Interface for Image Editing}.
\newblock In {\em Proc. of CHI}, pp. 2185--2194. {ACM}, 2013.

\bibitem{haotian}
H.~Li, M.~Xu, Y.~Wang, H.~Wei, and H.~Qu.
\newblock {A Visual Analytics Approach to Facilitate the Proctoring of Online
  Exams}.
\newblock In {\em Proc. of {CHI}}, pp. 682:1--682:17. {ACM}, 2021.

\bibitem{lin2020sportsxr}
T.~Lin, Y.~Yang, J.~Beyer, and H.~Pfister.
\newblock {SportsXR--Immersive Analytics in Sports}.
\newblock {\em ArXiv}, 2020.

\bibitem{narrativeorder}
N.~Montfort.
\newblock {Ordering Events in Interactive Fiction Narratives}.
\newblock In {\em Proc. of {Intelligent Narrative Technologies}}, pp. 87--94.
  {AAAI}, 2007.

\bibitem{draco}
D.~Moritz, C.~Wang, G.~L. Nelson, H.~Lin, A.~M. Smith, B.~Howe, and J.~Heer.
\newblock {Formalizing Visualization Design Knowledge as Constraints:
  Actionable and Extensible Models in Draco}.
\newblock {\em {IEEE} TVCG}, 25(1):438--448, 2019.

\bibitem{OffscreenCanvas}
Mozilla.
\newblock Offscreencanvas.
\newblock
  \url{https://developer.mozilla.org/en-US/docs/Web/API/OffscreenCanvas}, 2020.

\bibitem{directVideo2}
C.~Nguyen, Y.~Niu, and F.~Liu.
\newblock {Direct Manipulation Video Navigation in 3D}.
\newblock In W.~E. Mackay, S.~A. Brewster, and S.~B{\o}dker, eds., {\em Proc.
  of {CHI}}, pp. 1169--1172. {ACM}, 2013.

\bibitem{nguyen2016}
D.~T. Nguyen, W.~Li, and P.~O. Ogunbona.
\newblock {Human detection from images and videos: {A} survey}.
\newblock {\em Pattern Recognition}, 51:148--175, 2016.

\bibitem{Causality}
J.~Pearl.
\newblock {\em {Causality: Models, Reasoning, and Inference}}.
\newblock Cambridge University Press, USA, 2000.

\bibitem{perin2018state}
C.~Perin, R.~Vuillemot, C.~D. Stolper, J.~T. Stasko, J.~Wood, and
  S.~Carpendale.
\newblock {State of the Art of Sports Data Visualization}.
\newblock In {\em Proc. of CGF}, vol.~37, pp. 663--686. Wiley Online Library,
  2018.

\bibitem{Redmon2015}
J.~Redmon, S.~K. Divvala, R.~B. Girshick, and A.~Farhadi.
\newblock {You Only Look Once: Unified, Real-Time Object Detection}.
\newblock In {\em {Proc.} CVPR}, pp. 779--788. IEEE, 2016.

\bibitem{Ren2015}
S.~Ren, K.~He, R.~Girshick, and J.~Sun.
\newblock {Faster R-CNN: Towards Real-Time Object Detection with Region
  Proposal Network}.
\newblock In {\em {Proc.} NIPS}, pp. 91--99, 2015.

\bibitem{narrativeVis}
E.~Segel and J.~Heer.
\newblock {Narrative Visualization: Telling Stories with Data}.
\newblock {\em {IEEE} TVCG}, 16(6):1139--1148, 2010.

\bibitem{shicalliope}
D.~Shi, X.~Xu, F.~Sun, Y.~Shi, and N.~Cao.
\newblock {Calliope: Automatic Visual Data Story Generation from a
  Spreadsheet}.
\newblock {\em {IEEE} TVCG}, 1(1):1--1, 2021.

\bibitem{shih2017survey}
H.-C. Shih.
\newblock {A Survey of Content-Aware Video Analysis for Sports}.
\newblock {\em {IEEE TCSVT}}, 28(5):1212--1231, 2017.

\bibitem{secondspectrum}
S.~Spectrum.
\newblock Second spectrum.
\newblock \url{http://secondspectrum.com/}, 2020.

\bibitem{stein2018}
M.~Stein, T.~Breitkreutz, J.~H{\"{a}}ussler, D.~Seebacher, C.~Niederberger,
  T.~Schreck, M.~Grossniklaus, D.~A. Keim, and H.~Janetzko.
\newblock {Revealing the Invisible: Visual Analytics and Explanatory
  Storytelling for Advanced Team Sport Analysis}.
\newblock In {\em Proc. of BDVA}, pp. 1--9. {IEEE}, 2018.

\bibitem{stein2017bring}
M.~Stein, H.~Janetzko, A.~Lamprecht, T.~Breitkreutz, P.~Zimmermann,
  B.~Goldl{\"u}cke, T.~Schreck, G.~Andrienko, M.~Grossniklaus, and D.~A. Keim.
\newblock {Bring It to the Pitch: Combining Video and Movement Data to Enhance
  Team Sport Analysis}.
\newblock {\em {IEEE} TVCG}, 24(1):13--22, 2017.

\bibitem{tang2020design}
T.~Tang, J.~Tang, J.~Hong, L.~Yu, P.~Ren, and Y.~Wu.
\newblock {Design Guidelines for Augmenting Short-form Videos Using Animated
  Data Visualizations}.
\newblock {\em Journal of Visualization}, pp. 1--14, 2020.

\bibitem{tfjs}
TensorFlow.
\newblock Tensorflow.js.
\newblock \url{https://www.tensorflow.org/js}, 2020.

\bibitem{bodypiex2}
Tensorflow.
\newblock Tfjs-bodypix2.
\newblock \url{https://github.com/tensorflow/tfjs-models/tree/master/body-pix},
  2020.

\bibitem{Thompson2020}
J.~Thompson, Z.~Liu, W.~Li, and J.~T. Stasko.
\newblock {Understanding the Design Space and Authoring Paradigms for Animated
  Data Graphics}.
\newblock {\em CGF}, 39(3):207--218, 2020.

\bibitem{vizrt}
Vizrt.
\newblock Viz libero.
\newblock \url{https://www.vizrt.com/products/viz-libero}, 2020.

\bibitem{ttnet}
R.~Voeikov, N.~Falaleev, and R.~Baikulov.
\newblock {TTNet: Real-Time Temporal and Spatial video Analysis of Table
  Tennis}.
\newblock In {\em Proc. of CVPR}, pp. 3866--3874. {IEEE}, 2020.

\bibitem{vue}
VUE.
\newblock Vue.
\newblock \url{https://vuevideo.net}, 2020.

\bibitem{DBLP:journals/tvcg/WangWCZZW21}
J.~Wang, J.~Wu, A.~Cao, Z.~Zhou, H.~Zhang, and Y.~Wu.
\newblock {Tac-Miner: Visual Tactic Mining for Multiple Table Tennis Matches}.
\newblock {\em {IEEE} TVCG}, 27(6):2770--2782, 2021. doi: {{%
10\hspace{.1pt}\discretionary{.}{%
}{.}\hspace{.4pt}1109\discretionary{/}{%
}{/}TVCG\hspace{.1pt}\discretionary{.}{%
}{.}\hspace{.4pt}2021\hspace{.1pt}\discretionary{.}{%
}{.}\hspace{.4pt}3074576}}


\bibitem{WangSZCXMZ20}
Y.~Wang, Z.~Sun, H.~Zhang, W.~Cui, K.~Xu, X.~Ma, and D.~Zhang.
\newblock {DataShot: Automatic Generation of Fact Sheets from Tabular Data}.
\newblock {\em {IEEE} TVCG}, 26(1):895--905, 2020.

\bibitem{DBLP:journals/tvcg/WengZDMBZXW21}
D.~Weng, C.~Zheng, Z.~Deng, M.~Ma, J.~Bao, Y.~Zheng, M.~Xu, and Y.~Wu.
\newblock {Towards Better Bus Networks: {A} Visual Analytics Approach}.
\newblock {\em {IEEE} TVCG}, 27(2):817--827, 2021. doi: {{%
10\hspace{.1pt}\discretionary{.}{%
}{.}\hspace{.4pt}1109\discretionary{/}{%
}{/}TVCG\hspace{.1pt}\discretionary{.}{%
}{.}\hspace{.4pt}2020\hspace{.1pt}\discretionary{.}{%
}{.}\hspace{.4pt}3030458}}


\bibitem{voyager}
K.~Wongsuphasawat, D.~Moritz, A.~Anand, J.~D. Mackinlay, B.~Howe, and J.~Heer.
\newblock {Voyager: Exploratory Analysis via Faceted Browsing of Visualization
  Recommendations}.
\newblock {\em {IEEE} TVCG}, 22(1):649--658, 2016.

\bibitem{aoyu}
A.~Wu and H.~Qu.
\newblock Multimodal analysis of video collections: Visual exploration of
  presentation techniques in {TED} talks.
\newblock {\em {IEEE} TVCG}, 26(7):2429--2442, 2020.

\bibitem{wu2017ittvis}
Y.~Wu, J.~Lan, X.~Shu, C.~Ji, K.~Zhao, J.~Wang, and H.~Zhang.
\newblock {iTTVis: Interactive Visualization of Table Tennis Data}.
\newblock {\em {IEEE} TVCG}, 24(1):709--718, 2017.

\bibitem{DBLP:journals/tits/WuWDBXWZDC21}
Y.~Wu, D.~Weng, Z.~Deng, J.~Bao, M.~Xu, Z.~Wang, Y.~Zheng, Z.~Ding, and
  W.~Chen.
\newblock {Towards Better Detection and Analysis of Massive Spatiotemporal
  Co-Occurrence Patterns}.
\newblock {\em {IEEE} TITS}, 22(6):3387--3402, 2021. doi: {{%
10\hspace{.1pt}\discretionary{.}{%
}{.}\hspace{.4pt}1109\discretionary{/}{%
}{/}TITS\hspace{.1pt}\discretionary{.}{%
}{.}\hspace{.4pt}2020\hspace{.1pt}\discretionary{.}{%
}{.}\hspace{.4pt}2983226}}


\bibitem{DBLP:journals/tvcg/XieDW21}
X.~Xie, F.~Du, and Y.~Wu.
\newblock {A Visual Analytics Approach for Exploratory Causal Analysis:
  Exploration, Validation, and Applications}.
\newblock {\em {IEEE} TVCG}, 27(2):1448--1458, 2021. doi: {{%
10\hspace{.1pt}\discretionary{.}{%
}{.}\hspace{.4pt}1109\discretionary{/}{%
}{/}TVCG\hspace{.1pt}\discretionary{.}{%
}{.}\hspace{.4pt}2020\hspace{.1pt}\discretionary{.}{%
}{.}\hspace{.4pt}3028957}}


\bibitem{DBLP:journals/tvcg/XieWLDCZCW21}
X.~Xie, J.~Wang, H.~Liang, D.~Deng, S.~Cheng, H.~Zhang, W.~Chen, and Y.~Wu.
\newblock {PassVizor: Toward Better Understanding of the Dynamics of Soccer
  Passes}.
\newblock {\em {IEEE} TVCG}, 27(2):1322--1331, 2021.

\bibitem{ShuttleSpace}
S.~Ye, C.~Zhu-Tian, X.~Chu, Y.~Wang, S.~Fu, L.~Shen, K.~Zhou, and Y.~Wu.
\newblock {ShuttleSpace: Exploring and Analyzing Movement Trajectory in
  Immersive Visualization}.
\newblock {\em {IEEE} TVCG}, 27(2):860--869, 2021.

\bibitem{EmoCue}
H.~Zeng, X.~Shu, Y.~Wang, Y.~Wang, L.~Zhang, T.~Pong, and H.~Qu.
\newblock Emotioncues: Emotion-oriented visual summarization of classroom
  videos.
\newblock {\em {IEEE} TVCG}, 27(7):3168--3181, 2021.

\bibitem{EmoCo}
H.~Zeng, X.~Wang, A.~Wu, Y.~Wang, Q.~Li, A.~Endert, and H.~Qu.
\newblock Emoco: Visual analysis of emotion coherence in presentation videos.
\newblock {\em {IEEE} TVCG}, 26(1):927--937, 2020.

\bibitem{zhang2019pan}
C.~Zhang, Y.~Zou, G.~Chen, and L.~Gan.
\newblock {PAN: Persistent Appearance Network with an Efficient Motion Cue for
  Fast Action Recognition}.
\newblock In {\em Proc. of MM}, pp. 500--509. ACM, 2019.

\bibitem{Zhi2019}
Q.~Zhi, S.~Lin, P.~T. Sukumar, and R.~A. Metoyer.
\newblock {GameViews: Understanding and Supporting Data-driven Sports
  Storytelling}.
\newblock In S.~A. Brewster, G.~Fitzpatrick, A.~L. Cox, and V.~Kostakos, eds.,
  {\em Proc. of {CHI}}, p. 269. {ACM}, 2019.

\bibitem{marvist}
C.~Zhu-Tian, Y.~Su, Y.~Wang, Q.~Wang, H.~Qu, and Y.~Wu.
\newblock {MARVisT: Authoring Glyph-Based Visualization in Mobile Augmented
  Reality}.
\newblock {\em {IEEE} TVCG}, 26(8):2645--2658, 2020.

\bibitem{DBLP:conf/chi/ChenTWBQ20}
C.~Zhu-Tian, W.~Tong, Q.~Wang, B.~Bach, and H.~Qu.
\newblock Augmenting static visualizations with paparvis designer.
\newblock In {\em Proc. of {CHI}}, pp. 1--12. {ACM}, 2020.

\bibitem{chentimeline}
C.~Zhu-Tian, Y.~Wang, Q.~Wang, Y.~Wang, and H.~Qu.
\newblock {Towards Automated Infographic Design: Deep Learning-based
  Auto-Extraction of Extensible Timeline}.
\newblock {\em {IEEE} TVCG}, 26(1):917--926, 2020.

\end{thebibliography}
\end{document}